\documentclass[format=acmsmall, review=false, screen=true]{acmart}

\usepackage[norelsize, titlenumbered, linesnumbered, lined,ruled,commentsnumbered]{algorithm2e}
\usepackage{graphicx, subfigure}
\usepackage{caption}
\usepackage{epstopdf}
\usepackage{mathtools}
\usepackage{amsmath}
\usepackage{hyperref}
\usepackage{nameref}
\usepackage{todonotes}
\usepackage{enumitem}
\usepackage{multirow} 
\usepackage{tabularx}
\usepackage{placeins}
\usepackage{csquotes}
\usepackage{afterpage}
\usepackage{paralist}
\usepackage{setspace}
\usepackage{array,makecell,booktabs}
\usepackage[export]{adjustbox}[2011/08/13]
\setcitestyle{numbers,sort&compress}

\renewcommand\footnotetextcopyrightpermission[1]{} 
\newcolumntype{M}[1]{>{\centering\arraybackslash}m{#1}}
\DeclareMathOperator*{\argmin}{argmin}

\renewcommand{\algorithmcfname}{ALGORITHM}

\newcommand{\figtwoscale}{.32}

\SetAlFnt{\small}
\SetAlCapFnt{\small}
\SetAlCapNameFnt{\small}
\SetAlCapHSkip{0pt}
\IncMargin{-\parindent}

\begin{document}

\title[Network Structure Inference, A Survey: Motivations, Methods, and Applications]{Network Structure Inference, A Survey: Motivations, Methods, and Applications}  



\title{Network Structure Inference, A Survey: Motivations, Methods, and Applications}
\author{Ivan Brugere}
\orcid{0000-0001-8913-1733}
\affiliation{%
  \institution{University of Illinois at Chicago}
  \city{Chicago}
  \state{IL}
  \postcode{60607}
  \country{USA}}
\email{ibruge2@uic.edu}
\author{Brian Gallagher}
\affiliation{%
  \institution{Lawrence Livermore National Laboratory}
  \city{Livermore}
  \state{CA}
  \postcode{94550}
  \country{USA}}
\author{Tanya Y. Berger-Wolf}
\orcid{0000-0001-7610-1412}
\affiliation{%
  \institution{University of Illinois at Chicago}
  \city{Chicago}
  \state{IL}
  \postcode{60607}
  \country{USA}}
\email{tanyabw@uic.edu}

\begin{abstract}
Networks represent relationships between entities in many complex systems, spanning from online social interactions to biological cell development and brain connectivity. In many cases, relationships between entities are unambiguously known: are two users `friends' in a social network? Do two researchers collaborate on a published paper? Do two road segments in a transportation system intersect? These are  directly observable in the system in question. In most cases, relationship between nodes are not directly observable and must be inferred: does one gene regulate the expression of another? Do two animals who physically co-locate have a social bond? Who infected whom in a disease outbreak in a population? 

Existing approaches for inferring networks from data are found across many application domains and use specialized knowledge to infer and measure the quality of inferred network for a specific task or hypothesis. However, current research lacks a rigorous methodology which employs standard statistical validation on inferred models. In this survey, we examine (1) how network representations are constructed from underlying data, (2) the variety of questions and tasks on these representations over several domains, and (3) validation strategies for measuring the inferred network's capability of answering questions on the system of interest. 

\end{abstract}

\maketitle


\section{Introduction}
\label{sec:intro}


Networks represent relationships between entities in many complex systems, spanning from online social interactions to biological cell development and brain connectivity. In many cases, relationships between entities are unambiguously known: are two users `friends' in a social network? Do two researchers collaborate on a published paper? Do two road segments in a transportation system intersect? These are directly observable in the system in question \cite{PhysRevE.79.061916}. In \emph{most} cases, however, relationships between entities are not directly observable and must be inferred: does one gene regulate the expression of another? Do two animals who physically co-locate have a social bond? Who infected whom in a disease outbreak in a population?  


Networks are mathematical representations ({\em i.e.,} models) used to answer these types of questions about data collected on individual entities. There is a broad range of the questions over application domains and a variety of ways in which networks are used to answer these questions. How do we know whether a particular network on data is the most \textit{useful} representation for answering a given question? What is the ``right'' network representation and how do we \textit{compare} the utility of many possible representations for our particular question? Finally, how can we measure whether networks are the \textit{correct} models to answer a particular question on the system of interest?

Various approaches that use specialized knowledge to infer and measure the quality of inferred network for a specific task or hypothesis are found across many application domains. The rigor of these methodologies varies across domains and currently there are no common best-practices for evaluating the quality of networks inferred from data. Such practices would measure some combination of general statistical properties, including model significance-testing and uniqueness, sensitivity (\emph{i.e.,} the change in quality relative to underlying data measurements and model parameters), and generalizability (\emph{i.e.,} network quality measured over multiple hypotheses or tasks).

In this survey, we examine: 
    1) how network representations are constructed from underlying data;
    2) the variety of questions and tasks on these representations over several domains; and
    3) validation strategies for measuring the inferred network's capability for answering questions on the system of interest.

\subsection{Motivation: Networks Represent Complex Relationships}
\label{subsec:why}

Networks are a natural choice of data representation across many domains. First, networks by design represent \emph{higher-order structure} emerging from dyadic relationships, which serve as units of further analysis. These structures include neighborhoods, ego-nets, communities/modules, and connected components. For example, in the computational biology domain, clusters and motifs often represent shared biological function. The mutual connectivity provides stronger evidence for the common function than individual pairwise relationships. Individual, dyadic, neighborhood, or aggregate population analysis may be the most appropriate depending on the question of interest.  

Second, networks represent heterogeneity among entities in the local network topology. Rather than analysis on population aggregates (\emph{e.g.,} histograms), networks allow local querying of complex, non-metric attribute and feature spaces. \textit{Homophily} yields correlation in node attributes among entities close in the network. However, these correlations also tend to be non-monotonic: the \emph{most similar} node to a query entity may be arbitrarily distant in the network. Due to this correlation, network topology often represents local clusters as overlapping, heterogeneous relationships. For example, a user's `friends' in an online social network often cluster into functional units: friends from work, school, from the user's hometown etc., where nodes within each cluster are correlated in some--often unknown--attribute. The effectiveness of simple heuristics such as counting common neighbors in the link prediction problem \cite{Liben-Nowell2007} shows the latent  information within social networks.

Third, networks are \emph{interpretable} models for further analysis and hypothesis generation. Researchers can visually explore small networks and examine relationships between nodes to compare against their knowledge-base and intuition from the domain. Furthermore, a shared vocabulary of descriptive network measures enables researchers to compare networks according to density, degree distribution, clustering coefficient, centralities, diameter, average path length, triangle counts \cite{Tsourakakis:2009:DCT:1557019.1557111, Itai1978}, and graphlet distributions \cite{Pržulj12122004}. Many higher-level network measures have also recently been developed including robustness \cite{doi:10.1137/1.9781611973440.37, Purohit:2014:FIC:2623330.2623701}, local information efficiency \cite{Babaei:2016:EIN:2835776.2835826}, and routing efficiency \cite{Watts1998, 7003999}. Using these shared measures, researchers can compare properties across classes of networks and generate hypotheses explaining similarities or differences. 

Finally, networks are \emph{common} models for data and can be re-used in multiple studies. The breadth of tools and support for network analysis allows researchers of various disciplines to apply sophisticated off-the-shelf analysis and visualization techniques, as well as easier storage, querying, and portability in graph databases. Researchers also have a common vernacular developed in the area of ``network science,'' across biology, physics, and computer science \cite{brandes_robins_mccranie_wasserman_2013}. 


A cautionary note, however: when inferring networks from data, researchers ought to consider whether higher-order structures are \textit{meaningful} and which descriptive measures are appropriate on the inferred network. Available tools and convenience can motivate researchers to translate their problem into a network formulation, whether or not a network is the best model for the question of interest. This survey aims to clarify \textit{when} network models are appropriate and how they are inferred.  

\subsection{Inferring Network Models from Data}
\label{subsec:construction}

\emph{Inferred networks} are a class of networks where the node and/or edge definitions are inferred from non-explicitly relational data. Work in network science often focuses on applications of \emph{explicit networks}: network representations where the meaning of nodes and edges is unambiguous and categorical. For example, in the Facebook network, two adjacent users are categorically ``friends.'' In weighted networks such as a road transportation network, nodes represent road intersections and edges represent the road segments between them. These edges are unambiguous, change with low frequency, and have strict spatial constraints for new edge appearance. Measuring weights (\emph{e.g.,} travel time) is a matter of sensing/measuring traffic over the network, but these measurements are constrained to adjacent nodes on the known topology. In general, estimating weights in explicit network applications is a matter of accurately measuring a \textit{known} relationship on a \textit{known} topology, rather than learning a \emph{hypothesized} relationship on an \textit{unknown} topology. 
    
Seemingly explicit networks often also have biases and arbitrary choices (\emph{e.g.,} thresholding, see: Section \ref{subsubsec:interaction}) in the collection or construction of the network. These choices are out of the control of researchers using these networks. Starting from underlying data and translating data to a network model enables researchers to measure and account for these biases to ensure the network model actually represents the appropriate characteristics of the system.

\subsection{Challenges}
Inferring networks from data presents several challenges: \begin{compactitem}
\item[--] \textbf{Noisy measurements}: The underlying relationships of interest within real datasets are often noisy and confounded by overlapping relationships at varying scales (\emph{e.g.,} temporal, spatial). Furthermore, such errors are propagated when measuring higher-order properties of the inferred network (\emph{e.g.,} path lengths,  triangle counts, network centralities, communities).
\item[--] \textbf{Lack of ground truth and/or model assumptions}: Determining whether a particular method accurately encodes the relationship of interest in the network requires: (1) ground truth data, (2) model assumptions (\emph{e.g.,} Exponential Random Graph Model), or (3) some stability assumption (\emph{e.g.,} predictability over time). In many instances, no such assumptions or data are available and researchers are left with an under-determined problem and ad-hoc tuning of network model parameters \cite{10.1371/journal.pone.0090481}.
\item[--] \textbf{Large parameter spaces and sensitivity}: Many network inference methods have thresholds or other parameters which define a large space of possible networks. Often this leads to ad-hoc selection or parameter-space sampling (see: Section \ref{subsubsec:interaction}). Although sensitivity analysis is routine for method parameters of the subsequent task on the network (\emph{e.g.,} prediction, classification), most evaluation methodologies do not incorporate inferred network structure(s) into the sensitivity analysis. 
\item[--] \textbf{Varying model complexity}: Descriptive interpretation of edges, paths, and communities is increasingly challenging with additional model complexity. For example, a social network defined via thresholding on who-calls-whom call record data is more interpretable than a linear regression on node feature vectors. Determining the validity and interpretability of these higher-order features is crucial to the many analyses that use them. 
\end{compactitem}

\begin{figure}
\centering{\includegraphics[width=.85\columnwidth]{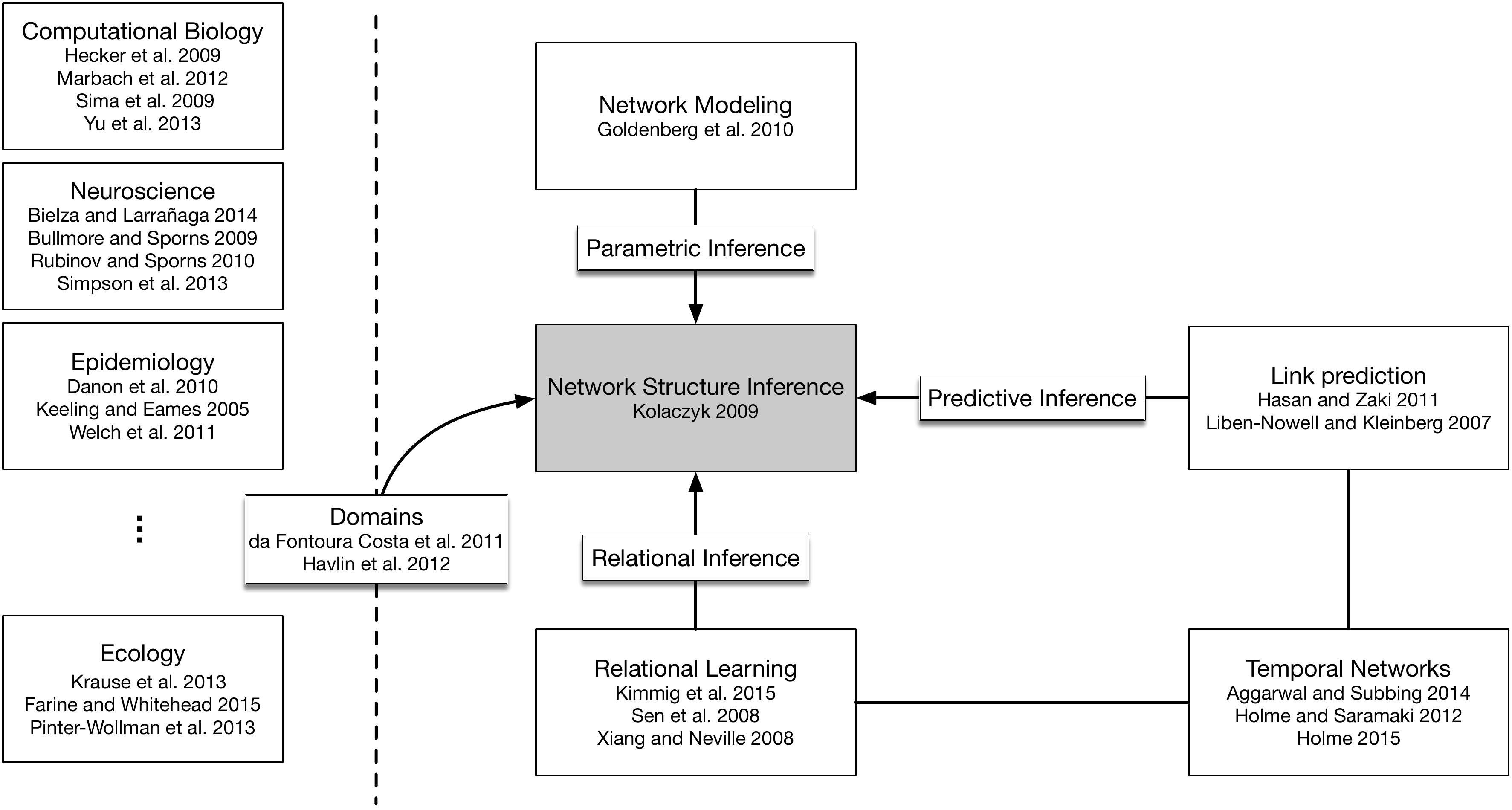}}
\caption{An overview of related areas in machine learning and network science for this survey, several surveyed domains, and principle surveys and introductory work for these sub-areas.}
\label{fig:metareview}
\end{figure}

\subsection{Outline of This Survey}

This survey organizes recent work focusing on inferring networks from data. We draw from several domains where these methods are applied and general method development from machine learning.
The study of inferring networks from data is largely unstructured. In Section \ref{sec:problem}, we first organize the work under a common problem formulation, taxonomy, and terminology. In Section \ref{sec:applications}, we examine various application domains to highlight unique challenges found in these areas.

\subsection{Audience}

Our survey is best suited for domain researchers using network inference in applications, as well as machine learning researchers interested in methods and methodology on networks. For novices in the networks space, Section \ref{sec:meta} structures many introductory surveys related to our work. 

Our survey is also relevant to the broader network science community working on traditional \textit{explicit} networks. These networks often represent an aggregation of longitudinal \textit{interactions} which are not informative to explain the \textit{current} behavior of the population. Closer scrutiny of edges in the network `as given' may yield more appropriate representations and affect all aspects of downstream analysis \cite{mlg2017_13}. Researchers routinely construct networks from `behavior' (\emph{i.e.,} node features) or other data to evaluate against their explicit network. Much of the work in scientific domains (including biology, ecology, chemistry) learn dynamic \emph{interaction} networks in the absence of ground truth. This work has developed varied strategies to evaluate the quality of networks in general for prediction, classification, or the discovery of robust edge relationships over time or question(s) of interest.

\subsection{Meta-Review: Comparison to Existing Surveys}
\label{sec:meta}
In this section, we provide a brief meta-review of existing surveys and work that is related but distinct from network structure inference. See Section \ref{subsec:taxonomy} for how these areas relate to our proposed taxonomy of related work within our problem.

Figure \ref{fig:metareview} provides a map of research in related but distinct network science problems. \citet{Kolaczyk2009} is the most relevant work to our survey. The author organizes related work in three categories: (1) the link prediction problem, where some edges and all nodes are known and the task is to infer new edges, (2) interaction networks where all nodes are known and the task is to infer edge relationships (\emph{e.g.,} by correlation), and (3) network tomography where some edges and nodes are known, and the task is to infer unobserved `interior' nodes and edges \cite{Ni:2010:EDR:1816288.1816298,4564479, Zhou2011}. Of these, (2) is primarily within the scope of what we define as the network structure inference; we focus with much greater depth on the network inference methodology and applications, where no edge definition is known a priori.  

We define network structure inference as distinct from the large body of work in relational learning. One branch contains attribute inference and prediction \emph{on} networks. Given a network, these methods infer missing attributes using local estimates \cite{sen2008collective, 4781232}, or link prediction: predicting edges at a later time-step or by node attribute similarity \cite{Liben-Nowell2007, Lü20111150, Hasan2011}. These are two fundamental tasks which serve as evaluation methodologies for many network inference methods. 

A second branch in relational learning focuses on inferring probabilistic relational models from data \cite{getoor20075, Kimmig2014}. While graphical models are one strategy applied to our problem, generally these models treat attributes or variables as nodes. Previous work in probabilistic relational models learns relationships between \emph{explicit} input and output graphs, using node attribute inference, entity resolution, and link prediction tasks in a \emph{supervised} setting \cite{namata:tkdd15}.

Other recent surveys cover broad statistical network modeling \cite{Goldenberg:2010:SSN:1734794.1734795} and multi-layer networks \cite{Boccaletti20141, Kivelä01092014}. Our survey draws on parametric network models, which is one class of model inferring parameters according to a particular distribution (\emph{e.g.,} attribute-edge joint distributions, \cite{Pfeiffer:2014:AGM:2566486.2567993}).  Research in network synthesis on multi-layer networks is one case of network structure inference for a particular task, where input data is a collection of networks.

Many network structure inference applications define edges by association measures over time (\emph{e.g.,} correlation in time series), so dynamics and temporal networks are an important aspect of network models for prediction tasks \cite{Holme201297, Aggarwal:2014:ENA:2620784.2601412, holme2015modern}. Finally, ``network representation learning'' is typically used to describe graph embedding methods. This is a rapidly growing area which aims to learn a low-dimensional representation of the network structure of a \textit{given} network \cite{DBLP:journals/corr/GoyalF17, 2017arXiv170905584H, 2017arXiv171108752C}, which preserves desired structural properties. These models are used for comparison of nodes or networks, or predictive tasks in the embedded representation. 

Our survey draws on several application areas. For a more exhaustive and general survey of complex networks in different application domains, \citet{doi:10.1080/00018732.2011.572452, Havlin2012} provide more breadth. However, we focus specifically on comparing motivations, evaluation methodologies, and challenges in several of these areas. Recent domain-focused meta-studies \cite{Marbach2012} and surveys in computational biology \cite{PMID:20190956, PMID:17981545, Hecker200986, Yu2013}, ecology \cite{Krause2013, Pinter-Wollman14062013,Proulx2005345, JANE:JANE12418}, neuroscience \cite{Bullmore2009, Rubinov2010, simpson2013, Bielza2014}, political science \cite{PSC:7968083}, and epidemiology \cite{Welch2011a,Keeling2005,Danon2011} all have significant discussion of network topology inference specific to the domain. However, few have network structure inference as a methodological focus and are limited to discussion of the single domain. Our survey focuses on challenges \emph{across} each of these areas. We provide value to domain researchers both within and across fields, as well as to researchers in machine learning interested in predictive model development on networks.

\section{Problem Description}
\label{sec:problem}

We give a high-level formulation of network structure inference to clarify the methodologies often encountered for this problem. This definition is very broad and not meant to be novel or exhaustive. We use this formalization for added precision of defining the constituent elements we will discuss. 

\subsection{Preliminaries}

We define a network $G = \langle \mathbf{V}, \mathbf{E}, \mathbf{A}\rangle$ as a tuple containing a set  $\mathbf{V}$ of $n$ nodes, $v_i \in \mathbf{V}$, a set $\mathbf{E}$ of $m$ node pairs, $e_{ij} \in \mathbf{E}$, and set $\mathbf{A}$ containing node and/or edge attribute sets. A particular attribute, the weight of an edge $w_{ij}$ is a scalar value, $|w_{ij}| \leq 1$, where $w_{ij} = 0$ denotes the absence of an edge. An unweighted network is a special case of a weighted network where $w_{ij} \in \{0, 1\}$. Edge and node \emph{features} are a particular type of attribute, derived by a function measuring some local edge or node property (\emph{e.g.,} node degree). Time-varying networks are defined as a $t$-length sequence of static network snapshots: $\mathbf{G} = (\mathbf{G}_1,...\mathbf{G}_k,...\mathbf{G}_t)$, with time-varying attributes and/or edges.\footnote{While there are other representations of time-varying (or temporal, dynamic) networks~\cite{Holme201297}, the time series of static networks is by far the most common.}

\subsection{Data Science Motivations for Network Structure Inference}


We organize network structure inference through the perspective of hypothesis-driven data science \cite{Cao:2017:DSC:3101309.3076253, Hey2009}. Under this perspective, the value of a network can be stated simply: for a question of interest on the underlying system and its relevant data, are networks an informative and useful \emph{data model} for \emph{better} answering the question?

\begin{figure}
\centering{\includegraphics[width=.85\columnwidth]{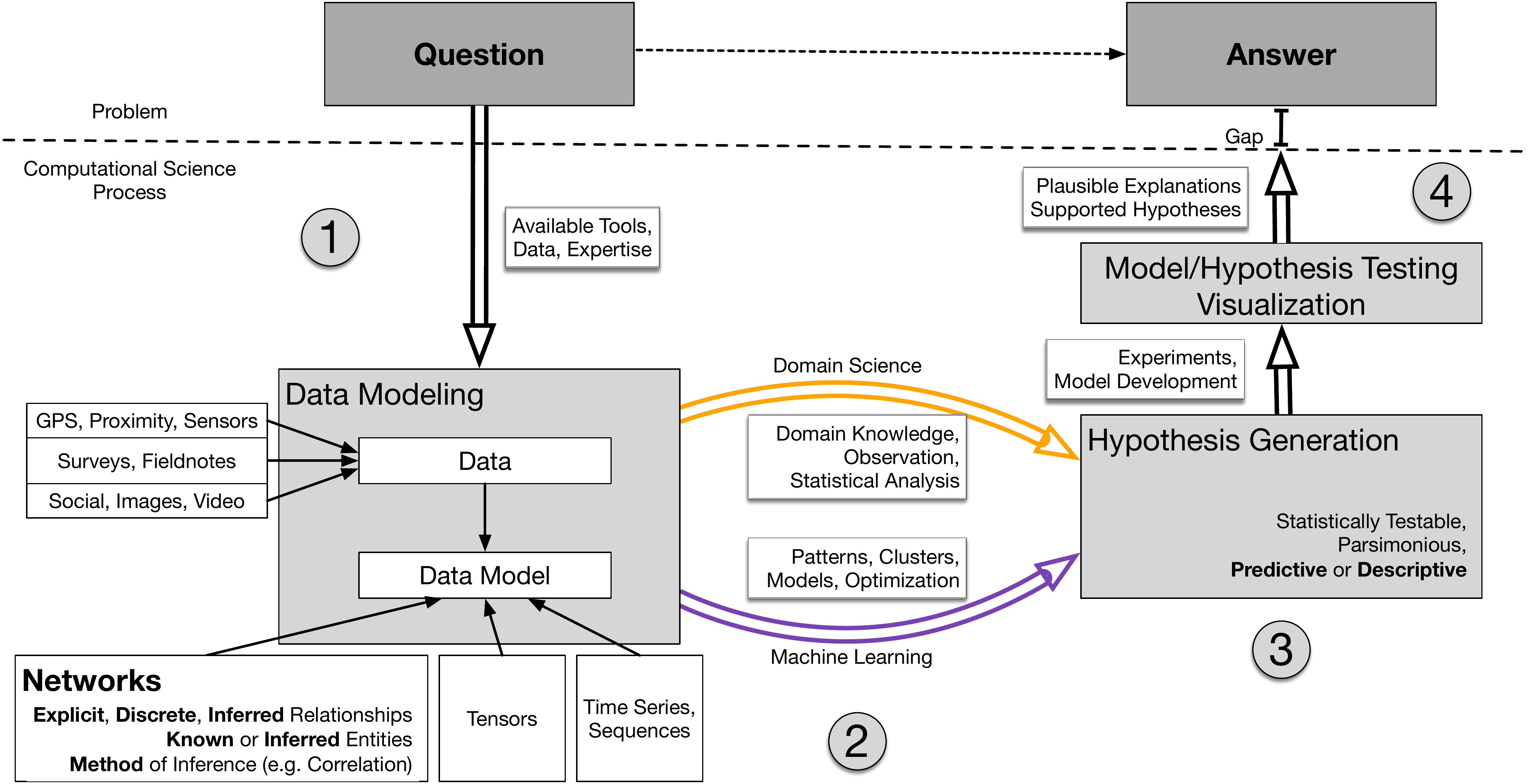}}
\caption{A common methodology for data science, with a focus on networks as models of underlying data. A researcher typically (1) uses \emph{relevant} data and \emph{appropriate} data models to formalize a particular question. (2) Traditional domain science (top, orange) typically combines direct observation and simple statistical models. Data-driven science with machine learning (bottom, purple) augments this intuition, hypotheses and statistical models with sophisticated features and patterns, yielding additional challenges for visualization, hypothesis testing and generation. (3) The output of this analysis is a plausible explanation or some supported hypothesis. (4) The result of data-driven science are considered to be shrinking this `gap' in understanding, toward robust and verifiable answers to the original question (see: \citet{Hey2009, Cao:2017:DSC:3101309.3076253}).}
\label{fig:datascience}
\end{figure}

Figure \ref{fig:datascience} gives an overview of a general data science methodology and introduces all the relevant terms to our taxonomy. We start with a broad question of interest and try to (1) locate or collect the data relevant to computationally modeling our question. There are several possible data sources, collection processes, and modalities for analysis, coupled with several possible representations (\emph{e.g.,} networks, tensors). Our choice of data representation informs and constrains our hypotheses about the question of interest. (2) Domain science (orange, top) and machine learning (purple, bottom) generate complimentary analyses for hypothesis generation. (3) Novel computational methods are developed to test these hypotheses. These tested models serve as a solution approaching an answer, (4) closing the gap in understanding for many questions in novel ways. For example, data-driven science has developed novel, large-scale methodologies relevant across scientific domains, including experimental design  \cite{Gui:2015:NAT:2736277.2741081,Backstrom:2011:NBT:1963405.1963492, Kohavi:2013:OCE:2487575.2488217} and randomization techniques \cite{efron93bootstrap, RSSB:RSSB12050}. 

Figure \ref{fig:datascience} highlights several levels of modeling which can impact the final capability for answering a question of interest. Often different individuals will be responsible for generating and collecting the underlying data, inferring or defining the network, and developing the predictive models. For example, machine learning researchers will rarely control the underlying sampling rate of specialized data collection processes in bioinformatics, neuroscience, or climate science. One novel approach in bioinformatics uses feedback from the network with active learning to guide the often costly and labor intensive data collection and experimentation process \cite{10.1371/journal.pcbi.1005466}.

In the machine learning community, there is a great deal of sensitivity analysis for the subsequent downstream predictive models on a network, without testing the parameters of the underlying network representation. This strategy is effective at producing sophisticated models ``given a network,'' but is less suited at answering our scientific questions on the original system. Previous work demonstrates that the incorrect network representation for a particular question can have a large impact on performance \cite{1708.06303}. Therefore the correct network representation may be \textit{more} important than developing the best predictive model.

\subsection{A Taxonomy for Network Structure Inference}
\label{subsec:taxonomy}

Below, we taxonomize work for the network inference problem according to: 

\begin{compactenum} 
\item The varying {\em difficulty} of defining edges and nodes, given the nature of the \emph{underlying data}.
\item The \emph{type of method} (\emph{e.g.,} mutual information, correlation) used for network inference.
\item The types of hypotheses and \emph{tasks} used to evaluate the quality of the network representation.
\end{compactenum}

\subsubsection{Explicit, Discrete, and Inferred Edge Definitions}

Networks model three broad types of relationships (\emph{i.e.,} edges), typically based on a measure of interaction between entities. Categorical, \emph{explicit} relationships are unambiguously known in the system--such as the `friend' or `follower' relations in online social networks. \emph{Discrete} interactions denote unambiguous transactions occurring between two entities--such as phone calls or text messages in mobile device data. The primary task of defining edges on these interactions is to select an appropriate threshold to measure the strength of the relationship. \emph{Inferred} interactions denote some statistical measure of similarity, beyond simple transaction counts. For example, the definition of a spatiotemporal \emph{co-location} interaction between two entities can be inferred by specifying ``how close'' for ``what duration,'' both of which are continuous thresholds. This is a more challenging to measure relative to discrete interactions.

\subsubsection{Explicit and Inferred Node Definitions}

Similarly, inferring nodes is of varying difficulty in different domains. In most applications, nodes are explicit in two ways: (1) the node definition is unambiguous (\emph{e.g.,} a `user', `animal', or `gene' is represented by a node) and (2) there is a node \emph{correspondence} between time-steps in the data (\emph{e.g.,} this particular node corresponds to a user `ivan' at each time step). Functional brain networks and climate networks are two examples of domains where the node definition is neither explicitly given nor is consistent over time. Here, nodes correspond to groups of two-dimensional earth surface pixels, or three-dimensional brain voxels. The \emph{scale} of node definition can dramatically change the performance of predictive models and the stability of descriptive statistics derived from the network \cite{Cammoun2012386}.

\subsubsection{Hypotheses and Tasks on Inferred Networks}


Figure \ref{fig:datascience} (2--3) illustrates the core novelty in machine learning and data science for hypothesis generation, model development, and testing. In machine learning, there is typically a bias toward evaluating models using a novel prediction tasks. However, typically these models do little to explain the underlying system and are more useful for generating new hypotheses. In the statistics domain and many natural sciences, there is a bias toward descriptive modeling, where the `task' is parameter inference on an assumed parametric (or non-parametric, but still with parameters, \cite{Wasserman:2006:NS:1202956}) statistical model on the data. These parameters are useful at interpreting the underlying system, such as fitting the exponent of degree distributions, or reporting other aggregate statistics on a network. Both of these aspects have a role for understanding the mechanisms of the underlying system.

Applications driven by descriptive models aim to reconstruct and describe the `hidden' relational structure with the greatest fidelity, relative to the domain knowledge base. A typical example is gene regulatory networks (GRNs). These networks are inferred on data measuring individual genetic expression over different experimental settings. In these networks, nodes represent genes or functional gene families and edges represent inferred positive or negative gene expression relationships (\emph{e.g.,} ``gene A reduces the expression of gene B under some context"). Inference of these networks typically identifies new, high quality \emph{candidate} regulation relationships given high accuracy of inferring known edges. These interactions can be experimentally tested to build greater understanding of cellular processes and to develop potential personalized medical treatments.  

In contrast, predictive methods aim to discover network structure which maximizes predictive performance and may not reconstruct the underlying process with the highest fidelity. Instead, they may focus on those aspects or modalities which are most predictive. Modeling the predictive aspects of the data allows researchers to learn regular relationships between modalities (\emph{e.g.,} call, text, and location in mobile phone data) or over time (\emph{e.g.,} periodicities). This informs further hypothesis generation to understand the \textit{mechanism} underlying the predictive relationships. However, often highly predictive relationships are also uninteresting and can drive the structure of the inferred network. Several domains such as climate science tend to remove known periodic dynamics as preprocessing of the underlying data, prior to inferring the network structure.

\subsubsection{Models for Network Structure Inference}

We organize related work broadly along the type of network structure inference \emph{model} used, including parametric, non-parametric, and thresholded interaction/correlation measures. In these groups, we categorize the type of task performed on the network, including edge and attribute prediction, descriptive analysis, and model selection. 

Table \ref{fig:citations} summarizes work across several domains, introducing an example scientific question driving the analysis, as well as the network structure inference model used to realize the network. In Table \ref{fig:citations} (Column `Model'), we label these models under two broad categories. 

First, \emph{parametric models} make assumptions on the \textit{distribution} of edges in the network. We identify graphical models (\textbf{GM}) and other network models using maximum likelihood methods (\textbf{ML}), relative to the assumption on the input data (\textit{e.g.}, the information flow between nodes \cite{Gomez-Rodriguez:2012:IND:2086737.2086741}). Causal models (\textbf{CM}) typically generate causal networks--a special case of graphical models--use Granger causality \cite{Granger1969} or other causal definitions \cite{Meek:1995:CIC:2074158.2074204, PMID:15360909}. These networks represent strong relationships between nodes which control for confounding factors of possible adjacent nodes. 

Second, \emph{non-parametric models} tend to directly measure interactions between nodes and use statistical tests to determine appropriate edge weights (see Section \ref{subsubsec:interaction}). We categorize work related to novel and `ad-hoc' interaction measures (\textbf{I}) between the data associated with pairs of nodes, correlation networks (\textbf{IC}) which measure linear, cross, or some other correlation, entropy (\textbf{IE}), frequency domain measures (\textbf{IF}), and regression (\textbf{R}). The \textit{edge definition} may be defined by an underlying parametric model (\textit{e.g.,} linear correlation), however we refer to these models as `non-parametric' because they don't have a model for the edge structure as a whole.

Table \ref{fig:citations} (Column `Task') categorizes rows within each domain by the type of task performed, under the caveat that one study may use several evaluation strategies, or that the actual task may be loosely described as the canonical task (\emph{e.g.,} edge prediction). 

First, we denote predictive tasks, including edge prediction (\textbf{PE}) and attribute prediction (\textbf{PA}). Attribute prediction can also describe prediction of the original data, or tasks such as change-point prediction \cite{mlg2017_17} and rank prediction, which are higher-order attribute tasks. Predictive models are relatively rare across application domains because researchers are interested in descriptive and interpretable models which give insight into the underlying system. We observe some specialization in both information networks and epidemiology, which aims to predict the extent or timing of an epidemic over a population in varying contact models. 

Second, descriptive analysis is broken by node-oriented statistics (\textbf{DN}), which includes reporting distributions of simple node statistics, such as degree distribution, clustering coefficient, correlation distributions, etc. This often constitutes the base-level exploratory analysis and qualitative evaluation of the inferred network. Role-oriented analysis (\textbf{DR}) aims to characterize nodes using network features by the structural roles they play in the system (\textit{e.g.,} bridges between social communities). This may also include factor analysis and interpretation. Other high-order analysis (\textbf{DH}) examines communities and larger subgraph structures beyond node and edge-based descriptive statistics. 

In descriptive and predictive cases, we find a sizable collection of work in model selection (\textbf{MS}). This work uses varying network representations with some evaluation criteria over models. These varying models correspond to different hypotheses to how the network may have been generated.    

\subsubsection{Evaluating Networks and Tasks}

Evaluation strategies over the entire methodology of Figure \ref{fig:datascience} measure the performance of the \emph{final task} (prediction or statistical inference). In this context, the network serves as a model of the data. However, the fidelity of this model in terms of fitting error, or reconstruction against a partial ground truth network does not measure the network's usefulness for answering questions on the original system. It is more appropriate to think of the space of networks as \emph{possible} representations with some utility for answering a specific question. `The network' is typically seen as uncovering the true relational structure of the data with some error \cite{Wang2012396}, rather than one of many representations for a particular purpose or question. 
Most scientific domains use an evaluation strategy focused on a final task (a network \emph{for} location prediction, brain activity response, etc.), because there is no evaluation framework for comparing the inferred network structure. For example, we simply cannot survey baboons \cite{Farine2016} as we can humans \cite{Eagle2009} to discover their real friendships. Furthermore, uncovering a general-purpose, robust network from complex data may not appropriately model the overlapping modalities of the data. For example, the existence of many \textit{contexts} may mean that functional brain networks are inherently probabilistic. Discovering the most \textit{general} network model will be less informative in any particular question because it does not account for this complexity \cite{1708.06303}. 

\subsection{Network Structure Inference}
\label{sec:def}

The network structure inference problem represents some input data as a network and \emph{validates} this network relative to performance on some task(s) or hypothesis. Often, model selection is done offline and we only see the result of the final network representation evaluated on the task. Currently, there exists no standardized methodology for evaluation of networks inferred from data. Our work structures this problem, particularly in the absence of network validation data.

The network topology inference problem combines two basic models: first, a \emph{network model} $\mathcal{R}(\mathbf{D}, \alpha) \rightarrow \mathbf{G}$ constructs network $\mathbf{G}$ on input data $\mathbf{D}$ under some parameters $\alpha$. This model may be parametric statistical models (\emph{e.g.,} Exponential Random Graph Models) or non-parametric and threshold-based interaction networks. Second, the problem uses a \emph{task model} $\mathcal{T}(\mathbf{G},  \beta) \rightarrow p_1,~p_2...$ on input $\mathbf{G}$ under some parameters, with task outputs (e.g. prediction `$p_i$'). These approximate the unknown ideal function $\mathcal{T}^*(G)$ of a network task (\emph{e.g.,} classification, prediction) with error $e()$. 

This formulation may seem overly simplistic. However, it succinctly clarifies the relationships between input data, the network model, and the task model. It explicitly formulates network $\mathbf{G}$ as a \emph{model} on data $\mathbf{D}$ for task $\mathcal{T}^*$, approximated by $\mathcal{T}$. This formulation captures simple \emph{interaction network} methodologies (see: Section \ref{subsubsec:interaction}) which separately infer the network (often by hand-tuned thresholds) and validate task performance, as well as parametric inference methods which learn the network model parameters. To our knowledge, all network structure inference methodologies can be broadly formulated in this pattern and all network inference models \emph{should} be formulated relative to a particular task or hypothesis. In much of the existing work, the network or task models will not be formulated explicitly, or possible model combinations may be sparsely-explored.

We can instantiate several tasks within this formulation. In the context of network prediction tasks, our predictive model can output predictions of (1) edges, (2) attributes, or (3) the original data. For one instantiation, on a validation edge-set $\mathbf{E}^*$, we can evaluate:
\small{} \begin{equation} 
	\label{eq:joint_opt} 
	\argmin_{\mathbf{G}} ~e(\mathcal{T}(\mathcal{R}(\mathbf{D}, \mathbf{\alpha}), \mathbf{\beta}), \mathbf{E}^*)
\end{equation}\normalsize{} This infers $\mathbf{G}$ over network model parameters $\alpha$ and task model parameters $\beta$. Practically, this might be done by a model selection over a grid search of parameter pairs \cite{1708.06303}, or an iterative re-training \cite{McAuley:2015:INS:2783258.2783381}. In this context, the suitability of both the network model, task model, and the appropriate error function will determine the performance of the inferred network. Network model selection  methodologies mitigate some of the biases from offline, hand-tuned network construction by exploring many possible model combinations.

This formulation highlights two key challenges in network structure inference. First, the parameter space of possible $\mathbf{G}$ from $\mathcal{R}(\mathbf{D}, \alpha)$ may be large and parameter search will likely be non-convex with respect to the performance of the task of interest. Second, well-performing models in the parameter-space may yield very different network topologies. Summarizing and reconciling these differences may be important for interpreting which network model is most suitable. This proliferation of plausible task and network models makes further hypothesis generation and testing more challenging for the researcher; more plausible models means additional interpretations of the mechanisms involved in the underlying behavior of the system.

\subsection{Network Models and Inference Methodologies}

Our problem description is flexible enough to incorporate varying types of network models (\emph{e.g.,} regression, correlation, parametric network models) and tasks (\emph{e.g.,} prediction, descriptive statistics and hypothesis testing). Below, we describe several formulations for network structure inference according to their network model and task model. 

\subsubsection{Interaction Networks}
\label{subsubsec:interaction}

The most prevalent class of network topology inference is measuring pairwise interactions (\emph{e.g.,} correlations) between nodes and choosing a threshold to define a sufficient degree of interaction. This threshold may be chosen by some statistical test, by tuning on some desired criteria (\emph{e.g.,} a desired network density), or ad-hoc offline testing. This has been discussed in the context of \emph{discrete} interactions as described above: 

\begin{displayquote}
``Inferring networks from pairwise interactions of cell-phone call or email records simply reduces down to selecting the right threshold $\tau$ such that an edge (u, v) is included in the network if u and v interacted more than $\tau$ times in the dataset. Similarly, inferring networks of interactions between proteins in a cell usually reduces to determining the right threshold.'' \cite{NIPS2010_4113}
\end{displayquote}


Researchers often make several application-specific decisions around these thresholds:

\begin{displayquote}
``From this complete correlation graph, only the edges with significant correlation ($>0.5$) were retained. But using the same threshold for positive and negative correlations is not appropriate as negative correlations are usually weaker and many nearby locations have high positive correlation'' \cite{Kawale2013}
\end{displayquote}

\begin{displayquote}
``We let $\delta$ as a user-controlled parameter, where larger $\delta$ values correspond to less predicted regulations, and only focus on designing a significance score $s(t,g)$ that leads to `good' prediction for some values of $\delta$'' \cite{23173819}
\end{displayquote}

These methods typically produce a fixed network model and explore networks under varying threshold settings in an offline trial-and-error fashion. Unfortunately, these negative results are rarely reported, so we often only have a description of the final network definition. 

We use our formulation for this process. Assume the interaction threshold $\tau$ is given by hand-tuning or domain knowledge; we have some feature matrix $\mathbf{D}$ that has some similarity between features, measured by $\mathcal{R}()$, and $\mathcal{T}()$ is an edge prediction task evaluated on $\mathbf{E}^*$. This method is expressed as: 
\small{}\begin{equation}\small
	\label{eq:simple_th} 
	\mathcal{R}(\mathbf{D}, \mathbf{\tau}) \rightarrow \mathbf{G};~ \argmin_{\mathbf{\beta}} ~e(\mathcal{T}(\mathbf{G}, \mathbf{\beta}), \mathbf{E}^*)
\end{equation}\normalsize{} When these interaction networks are evaluated in the absence of ground truth, the network may be measured through autocorrelation. In this case, the same network inference is applied, $\mathcal{R}(\mathbf{D^*}, \mathbf{\alpha})  \rightarrow \mathbf{G^*}$ for some hold-out data $D^*$.

\subsubsection{Parametric Network Models and Parameter Inference}

Maximum-likelihood methods assume some parametric model family to represent relationships between nodes, such as time between interaction, likelihood of information transmission over time. For clarity, we work through one specific application, in epidemiology and information networks such as blogs, although the pattern is similar in other applications. Structure inference methods on disease transmission networks share the assumption that we are observing the `arrival' of infection or information/attribute value at nodes (\emph{i.e.,} computers, blog pages, individuals) over time, but are unable to observe the topology which transmitted the information. For an edge $e_{ij}$, the likelihood of information transmission (or infection, in the epidemiology context) time difference $t_j - t_i$ is given by an information transmission/infection model, yielding the likelihood that $v_i$ transmitted information to $v_j$ \cite{NIPS2010_4113}.

Where input data $\mathbf{D}$ are infection times of each node, we can formulate these methods as: \small{}\begin{equation} 
	\label{eq:maximum_likelihood} 
	\argmin_{\mathbf{\alpha}} ~\mathcal{R}(\mathbf{D}, \mathbf{\alpha}) \rightarrow \mathbf{G};  ~\argmin_{\mathbf{\beta}} ~e(\mathcal{T}(\mathbf{G}, \mathbf{\beta}), \mathbf{E}^*)
\end{equation}\normalsize{} In information network applications, $\mathbf{E}^*$ is typically provided by a known network. Processes are simulated on this network to generate input data for the maximum-likelihood relational data model $\mathcal{R}()$. This method is used to `reconstruct' $\mathbf{E}^*$ only from input data $\mathbf{D}$. 

\subsubsection{Learning Network Structure and Task Model Parameters}
\label{subsubsec:joint}
Previous work in statistical relational learning on \emph{explicit} networks has focused on learning relationships between (categorical) attributes and a predictive task, such as link prediction \cite{Gong:2014:JLP:2611448.2594455,namata:tkdd15}, and distinguishing correlated effects between these processes \cite{LaFond:2010:RTD:1772690.1772752}. 

Previous work has also used a methodology for the network structure inference problem which maximizes performance of particular task(s) on the inferred network \cite{DeChoudhury:2010:IRS:1772690.1772722, Farine2016}. Consider interaction networks over varying thresholds $\tau$. A naive solution for this type of approach is to explore the parameter space of $\mathcal{R}(\mathbf{D}, \tau)$ and evaluate the task performance on each inferred network. However, this is very costly, especially as $\mathcal{R}()$ requires more parameters. 

Learning network structure paired with task parameters $\mathcal{T}()$ yields both the network and task definition according to some optimization or model selection criteria. For example, previous work has learned network models that iteratively discover features in a supervised LDA model, and the logistic regression weights that perform well for an edge prediction task \cite{McAuley:2015:INS:2783258.2783381}. Other work uses model selection to select task and network models by performance \cite{2017arXiv170507967V, 1708.06303}, or minimum description length (MDL) \textit{encoding cost} \cite{2017arXiv171005207B}. A related area of work treats dynamic networks as a sampling process over varying sampling \textit{rates} \cite{Ribeiro2013}, where network models are evaluated according to the most predictive representation for the given task \cite{mlg2017_17, Sulo:2010:MST:1830252.1830269}.

In Equation \ref{eq:joint_opt}, we have already formulated paired evaluation of network and task parameters:
\small{}
\begin{equation} 
	\label{eq:joint_opt2} 
	\argmin_{\mathbf{G}} ~e(\mathcal{T}(\mathcal{R}(\mathbf{D}, \mathbf{\alpha}), \mathbf{\beta}), \mathbf{E}^*)
\end{equation} \normalsize{}With more context we now see $\mathcal{R}()$ can construct networks of varying interaction measures, sampling rate criteria, etc., and this methodology selects the $\mathbf{G}()$ corresponding to minimum error on $\mathcal{T}()$. Therefore this simple formulation defines a very general model selection methodology.

\section{Applications of Network Structure Inference} 
\label{sec:applications}

Applications of network structure inference are too numerous to  meaningfully cover. Instead, we look closely at a few applications where inferring networks from data is illustrative of the challenges and expertise in these domains.

\renewcommand{\arraystretch}{1.02}
\newcommand\cwidthtask{2cm}
\noindent\begin{table*}
\resizebox{.80\textwidth}{!}{
\centering
\small
\begin{tabularx}{1.0\textwidth}{m{\cwidthtask} m{\cwidthtask} >{\centering\arraybackslash}m{.6cm} |>{\centering\arraybackslash}m{.725cm}|>{\centering\arraybackslash}m{1.1cm}|>{\centering\arraybackslash}m{5cm}}
	 
	  \cmidrule[3pt]{1-6}
	  {\em Domain} & {\em Example} & {\em Sec.} & {\em Model} & {\em Task} & {\em Citations}\\
	  \cmidrule[1pt]{1-6}
	  \multirow{5}{\cwidthtask}{Computational Biology} & \multirow{5}{\cwidthtask}{Discover interactions between genes in cellular processes} & \multirow{5}{*}{\centering\ref{subsec:grn}} & \textbf{I} & \multirow{5}{*}[-1.5em]{\textbf{DH}, \textbf{DR}}& {\smaller \citet{ZhangHorvath2005}} \\ 
	  \cline{4-4} \cline{6-6}
	  & & & \textbf{IE} & & {\smaller \citet{Faith2007, Butte00mutualinformation, Meyer2007}} \\
	  \cline{4-4} \cline{6-6}
	  & & & \textbf{GM} & & {\smaller \citet{Toh2002, Mukherjee2008, 20860793, PMID:15360909, Allen:2012:LGM:2468872.2468957}} \\
	  \cline{4-4} \cline{6-6}	
	  & & & \textbf{R} & & {\smaller\citet{Yuan2006, 23173819}} \\	
	  \cline{4-6}
	  & & & \textbf{CM} & \textbf{PE} & {\smaller \citet{Feizi2013, Barzel2013}}  \\	  
	  \cline{1-6}
	  \multirow{3}{\cwidthtask}{Climate} & \multirow{5}{\cwidthtask}{Describe relationships in environmental system dynamics} & \multirow{5}{*}{\centering\ref{subsec:climate}} & \textbf{IC}  & \multirow{2}{*}[-.3em]{\textbf{DN}, \textbf{PA}} &  {\smaller\citet{Tsonis2004497, PhysRevLett.100.228501, SAM:SAM10100, npg-18-751-2011, Kawale:2012:TSS:2339530.2339634, 10.1371/journal.pone.0153703}} \\	  
	  \cline{4-4} \cline{6-6}
	  & & & \textbf{IE} & & {\smaller\citet{0295-5075-87-4-48007, e15062023}}  \\	
	  \cline{4-6}
	  & & & \textbf{CM} & \multirow{3}{*}{\textbf{DN}} & {\smaller\citet{Ebert-Uphoff2012, Kretschmer2016, Runge2013}}  \\
	  \cline{4-4} \cline{6-6}
	  & & & \textbf{GM} & & {\smaller\citet{:/content/aip/journal/chaos/24/2/10.1063/1.4870402}}  \\
	  \cline{4-4} \cline{6-6}
	  & & & \textbf{R} & & {\smaller\citet{PhysRevLett.115.268501}}  \\	
	  \cline{1-6}	
	  \multirow{4}{\cwidthtask}{Neuroscience} & \multirow{4}{\cwidthtask}{Model relationships between brain regions, physiological structure, and function} & \multirow{4}{*}{\centering\ref{subsec:brain}} & \textbf{IC}  & \multirow{3}{*}[-2.5em]{\textbf{DN}, \textbf{MS}} &  {\smaller\citet{Bialonski2011, Zalesky20122096}}  \\		  
	  \cline{4-4} \cline{6-6}
	  & & & \textbf{IF} & & {\smaller\citet{Lachaux2002, Zhan2006, Ponten2016, PMID:10600019}}  \\	
	  \cline{4-4} \cline{6-6}
	  & & & \textbf{CM} & & {\smaller\citet{Ramsey20101545, Roebroeck2005230,Dhamala2008354,Rosa201266, Friston2011, David2008}}\\
	  \cline{4-6}
	  & & & \textbf{ML} & \textbf{PA} & {\smaller\citet{Papalexakis:2014:GBM:2623330.2623639, doi:10.1137/1.9781611974973.22}}  \\
	  \cline{1-6}	
	  \multirow{3}{\cwidthtask}{Epidemiology} & \multirow{3}{\cwidthtask}{Model hidden networks from observed infections} & \multirow{3}{*}{\centering\ref{subsec:information}} & \textbf{I} & \multirow{3}{*}{\textbf{PA}, \textbf{MS}} & {\smaller\citet{Haydon2003, 1517844}} \\
	  \cline{4-4} \cline{6-6}
	  & & & \textbf{GM} & & {\smaller\citet{BRITTON2002,GROENDYKE2011,Stack06012013} }\\
	  \cline{4-4} \cline{6-6}
	  & & & \textbf{ML} & & {\smaller\citet{NIPS2010_4113, Gomez-Rodriguez:2012:IND:2086737.2086741, ICML2011Gomez_354, NIPS2012_4582, Netrapalli:2012:LGE:2318857.2254783}} \\
	  \cline{1-6}
	  \multirow{4}{\cwidthtask}{Ecology} & \multirow{4}{\cwidthtask}{Describe and predict animal behavior} & \multirow{2}{*}{\centering\ref{subsec:ecology}} & \textbf{I} & \multirow{2}{*}{\textbf{DH}, \textbf{MS}} & {\smaller\citet{Psorakis07112012, Haddadi2011, Aplinrspb20121591, MEE3:MEE312553} }\\	  
	  \cline{4-4} \cline{6-6}
	  & & & \textbf{IE} & & {\smaller\citet{Barrett2108} }\\
	  \cline{4-6}
	  & & & \multirow{2}{*}{\textbf{R}} & \textbf{DN} & {\smaller\citet{MEE3:MEE312383, MEE3:MEE312772}} \\ 
	  \cline{5-6}
	  & & & & \textbf{PE} & {\smaller\citet{Farine2016,Fletcher29112011}} \\ 	  
	  
	  \cline{1-6}	
	  Mobile & Predict social influence on individual mobility  & \ref{subsec:mobility} & \textbf{I} & \multirow{1}{*}{\textbf{PE}, \textbf{DN}} & {\smaller\citet{Sekara2014, Mastrandrea2015, Eagle2009, DeChoudhury:2010:IRS:1772690.1772722}} \\	  
	  \cmidrule[1.0pt]{1-6}	
	  \multicolumn{6}{c}{\emph{Interaction}, \textbf{IC}: Correlation, \textbf{IE}: Entropy, \textbf{IF}: Frequency, \textbf{I}: Novel measures}\\
	  \multicolumn{6}{c}{\textbf{CM}: Causal model \textbf{GM}: Graphical model, \textbf{ML}: Maximum likelihood, \textbf{R}: Regression}\\	  
	  \cmidrule[1.0pt]{1-6}
	  \multicolumn{6}{c}{\emph{Prediction}, \textbf{PA}: Attributes, \textbf{PE}: Edges}\\
	  \multicolumn{6}{c}{\emph{Descriptive Analysis}, \textbf{DN}: Nodes, \textbf{DR}: Roles, \textbf{DH}: Other high-order, \textbf{MS}: Model Selection}\\
	  	
	\cmidrule[3pt]{1-6}
\end{tabularx}}


\caption{A summary of related work, across domains}
\label{fig:citations}

\end{table*}

\subsection{Computational Biology: Discovering New Genetic Regulation}
\label{subsec:grn}
\begin{figure}
\centering{\includegraphics[width=.75\columnwidth]{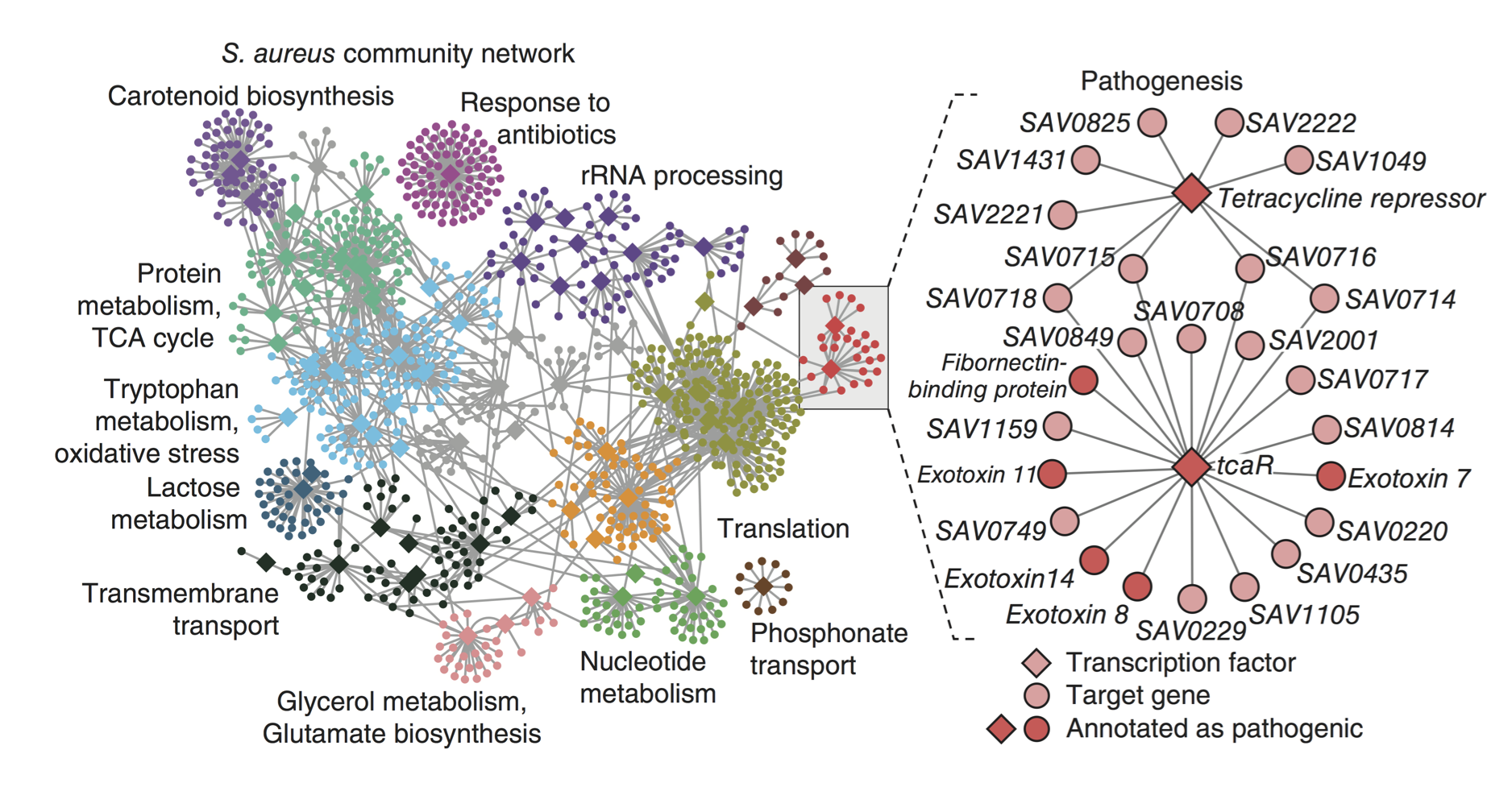}}
\caption{A consensus gene regulatory network (GRN) of \emph{Staphylococcus aureus} generated from microarray data in the DREAM5 challenge ensemble methodology (see: Figure \ref{fig:dream5}). Network modules in agreement with Gene Ontology (GO) database are colored and labeled according to their shared function (grey are not coherent in GO). The detail (right) shows inferred gene regulation related to pathogenesis. (From \citet{Marbach2012}).}
\label{fig:cell}
\end{figure}

Networks are constructed in computational biology to model many different processes. Protein-protein interaction (PPI) is the most common type of network in this domain, constructed mostly experimentally through, for example, yeast two-hybrid experiments others that physically test for binding of one protein to another or other types of physical interactions. Metabolic networks are process networks that model relationships between enzymes, metabolites (typically nodes) on processes (typically edges) such as reactions or pathways. We focus only on gene regulation networks (GRNs) and gene co-expression networks (GCNs), which infer relationships between genes under different experimental scenarios (for further reading, see: \cite{PMID:20190956}). 

\subsubsection{Underlying Data in Gene Regulatory Networks}

Next-generation high-throughput microarray technologies allow the sequencing of genomes and measuring the \emph{expression} of particular genes at a large scale and low cost \cite{Shendure2008}.  `Expression' is the measurement of how groups of genes produce different phenotypic specializations through the production of different proteins. In cellular development, different gene co-expression can be responsible for RNA translation or nucleotide metabolism, yielding many complex functions from gene interaction (see: \cite{Barrett01012013}). Figure \ref{fig:cell} illustrates a small network with annotated functional clusters.\footnote{All figures reprinted with permission and attribution.} 

The output of microarray analysis (with notable simplification) is a 2D data matrix of numeric values measuring the expression of a gene (row), on a particular experimental design, subject, or time step (column) \cite{Bar-Joseph2012}. Defining edges between genes simplifies to comparing expression profiles across the different columns of the data.

\subsubsection{Discovering New Gene Interactions From Data}

The high-level `task' for gene regulatory networks is \emph{link prediction} on the network learned from data, to discover unknown gene regulation candidates which can be experimentally tested. The network model inferred from data should agree with databases of biologically-known interactions and function, while providing few verified false-positive regulations. 

Table \ref{fig:citations} illustrates that the inference of GRNs is very mature relative to other domains, since the inferred network is clearly verifiable according to current domain knowledge, and directs future investigation and hypothesis generation. One unique challenge in gene regulation is the issue of confounding factors, including indirect and transitive associations, resulting in many spurious edges. Recent work has measured these `direct' (\emph{e.g.,} causal) edges in noisy expression datasets \cite{Barzel2013, Feizi2013}.

\begin{figure}
\centering{\includegraphics[width=.75\columnwidth]{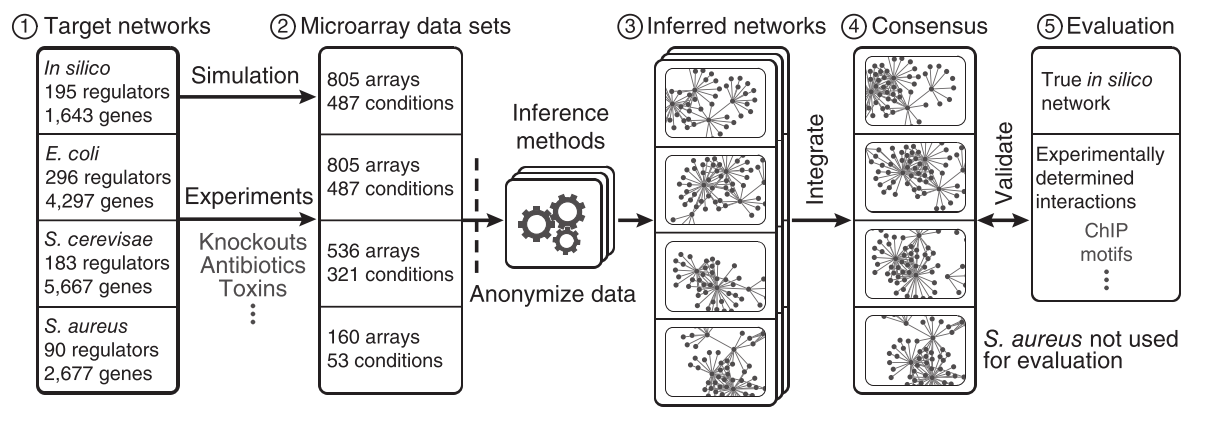}}
\caption{The general methodology design for the DREAM5 network inference challenge. This methodology (1-2) generates one simulation dataset and three experimental microarray datasets from three well-studied species. The 29 participating inference methods all (3) generate inferred network output, (4) a consensus network is constructed for each dataset, and then (5) validated against known edges in synthetic networks and against experimentally known edges in two of the real datasets. (From \citet{Marbach2012}).}
\label{fig:dream5}
\end{figure}

\subsubsection{Translating Gene Expression Data to Interaction Discovery}

\citet{Marbach2012} introduce an ensemble approach associated with the DREAM5 network inference challenge. The authors present 29 different network inference methods across different model types, including regression, mutual information, correlation (a.k.a. `relevance networks' in this domain), Bayesian networks, ensembles, and other novel approaches (\emph{e.g.,} random forests, neural networks, Gaussian mixture models, etc.) This list demonstrates the maturity and variety of methods applied to this problem. These methods use pairwise (\emph{e.g.,} gene-to-gene correlation) or group-wise (\emph{e.g.,} many-to-one group LASSO) measures of interaction intensity, which yield a directed, unweighted network signifying ``gene A regulates gene B.'' 

Figure \ref{fig:dream5} illustrates an ensemble methodology for the DREAM5 network structure inference challenge. The authors (1) generate a ground truth network on three species using experimental trials as well as synthetic network data. In (2), these experiments yield four different datasets of biologically-tested networks of gene regulation, as well as the raw (non-relational) gene expression. The collection of inference methods produce (3) inferred networks on each of the four datasets. The authors (4) integrate these 29 different inferred networks to produce an ensemble network. This network is (5) validated against the ground-truth networks generated in step (1). Finally, the authors show that the ensemble method discovered $59$ potential interactions, of which $29$ show some support and $20$ show strong support for being biologically significant. 

We examine this process in such detail to demonstrate the `complete' methodology for network structure inference, from input data, to network representation, to a final output (learning the regulatory network). Within our formulation, this methodology does not have an explicit \emph{task model}, as the network itself is the object of interest. Therefore, this application is typically \emph{descriptive} modeling against evaluation data. Although these models do `predict' new potential gene regulation via unknown edges, these interactions are usually checked manually through experimentation.

\subsection{Environmental Science: Discovering Climate Relationships and Predicting Outcomes}
\label{subsec:climate}
Networks inferred to understand climate dynamics are among the most difficult to model of any domain, with much of the work to formalize and validate these networks still in early development. Within this domain, researchers want to discover robust, \emph{causal} relationships between climatic variables, over different spatial regions of earth. This modeling can improve prediction of changing hydrological processes, land-cover, ecosystem productivity, and polar or sea ice cover, which are key aspects for climate change mitigation. Two unique challenges exist for inferring climate networks:

\begin{compactenum}
\item Input data is typically noisy, highly spatially-autocorrelated, multivariate time series of climatic variables collected over varying regimes and sensor (\emph{e.g.,} satellite) quality. Domain scientists produce a consistent``reanalysis'' data products which attempt to mitigate problems of varying sensors and errors in data. However, inferring accurate networks from these data requires significant understanding of the data ingestion process \cite{doi:10.1175/JCLI-D-12-00823.1, Levitus2013} and its introduced biases and variability.
\item The structure of climate networks is not well understood aside from a handful of climate indices--coarse spatial locations on earth where dynamics are well-studied and regulate or correlate with other environmental processes (\emph{e.g.,} El Ni\~{n}o and La Ni\~{n}a oscillation cycles). Therefore, validating correctness of the inferred network is suitable for unsupervised strategies such as relational or predictive modeling of the original data. 
\end{compactenum}

\subsubsection{Network Inference Methods in Climate and Environmental Science}

Nearly all studies constructing climate networks use some time series similarity as an underlying relational measurement. Previous work has used linear correlation \cite{Tsonis2004497, Donges2009, SAM:SAM10100, PhysRevLett.100.228501} or mutual information \cite{Donges2009, e15062023},  either a hand-picked \cite{Donges2009, Tsonis2004497} or a simple statistical test \cite{PhysRevLett.100.228501} to set similarity threshold $\tau$--where similarity greater than $\tau$ is considered a binary edge in the network. 

There is considerable focus on formulating these simple pairwise comparison methods and often the `recipe' of the network, according to parameter settings and preprocessing choices, varies greatly from study to study. These networks are typically binary rather than weighted, because the final output of interest is a binary decision on the similarity distribution: (\emph{e.g.,} [``significant'', ``not significant'']). However, typically these measures will have no `natural' threshold which gives this binary classification. Instead, these networks can be gradually densified or sparsified by weakening or strengthening the similarity threshold. 

Descriptive statistical work has been very popular \emph{downstream} from these varying network `recipes.' \citet{Donges2009} reports clustering coefficient, betweenness centrality, closeness centrality on correlation and mutual information networks (Equation \ref{eq:corr}--\ref{eq:MI}). \citet{Tsonis2011} reports community structure which tends to cluster in spatially-contiguous locations, on account of the autocorrelation present in these networks. Little work focuses on evaluating these networks as predictive models for the input data. \citet{SAM:SAM10100} use both descriptive statistics and predictive performance to evaluate the inferred network. 

As work utilizing network models grows in this climate and environmental science, researchers  have developed more sophisticated techniques for determining edge significance \cite{Kawale:2012:TSS:2339530.2339634}, or conditional dependencies using causality \cite{Ebert-Uphoff2012, Kretschmer2016, Runge2013}. 

\subsubsection{A Methodology for Translating Environmental Sensing Data to Interaction Network}

We will step through a concrete example of constructing a climate network from spatially-gridded time series data of global surface air temperature (SAT) \cite{Donges2009}. The authors measure the similar dynamics of pairwise earth locations (corresponding to nodes $n_i, n_j$) and construct edges between locations with `significant' similarity. The authors use two standard measures, linear correlation and mutual information between time series $X_i$ and $X_j$:\small{}
\begin{equation} \label{eq:corr}
P_{ij} = \frac{\sum\limits_{t=1}^{|X_i|}(X_{i,t} - \bar{X_i})(X_{j,t} - \bar{X_j})}{\sqrt{\sum\limits_{t=1}^{|X_i|}(X_{i,t} - \bar{X_i})^2}\sqrt{\sum\limits_{t=1}^{|X_j|}(X_{j,t} - \bar{X_j})^2}}
\end{equation}

\begin{equation} \label{eq:MI}
M_{ij} = \sum\limits_{b=1}^{|B|} p_b(X_i, X_j) \text{log}\frac{p_b(X_i, X_j)}{p_b(X_i)p_b(X_j)}
\end{equation}
\normalsize{} Equation \ref{eq:corr} is the sample Pearson correlation between two time series, where $\bar{X_i}$ is the sample mean of time series $X_i$. The denominator represents the product of the sample standard deviations of $X_i$ and $X_j$. This measures linear relationship of $X_i$ and $X_j$ over the length of the time series. Equation \ref{eq:MI} is the discrete mutual information estimate between two time series, where $p_b(X_i, X_j)$ is the joint cumulative distribution of the $b$-th discretization window, and $p_b(X)$ is the marginal cumulative distribution of the $b$-th discretization window. This measure compares the shape of the joint and marginal probability density functions. When the joint distribution is equal to the product of marginal distributions: $\text{log}(1) = 0$ yields no mutual information  (\emph{e.g.,} $X_i$ and $X_j$ are independent). 

Varying similarity threshold $\tau$ produces networks of varying edge densities $\rho$ and other network measures. The authors select thresholds for correlation and mutual information ($\tau_{corr} = 0.682$ and $\tau_{MI} = 0.398$) such that they produce the same network density ($\rho= 0.005$). 

\begin{figure}
	\centering     
	\subfigure[$P_{ij}$ vs. distance]{\label{fig:clim_a}\includegraphics[width=.25\columnwidth]{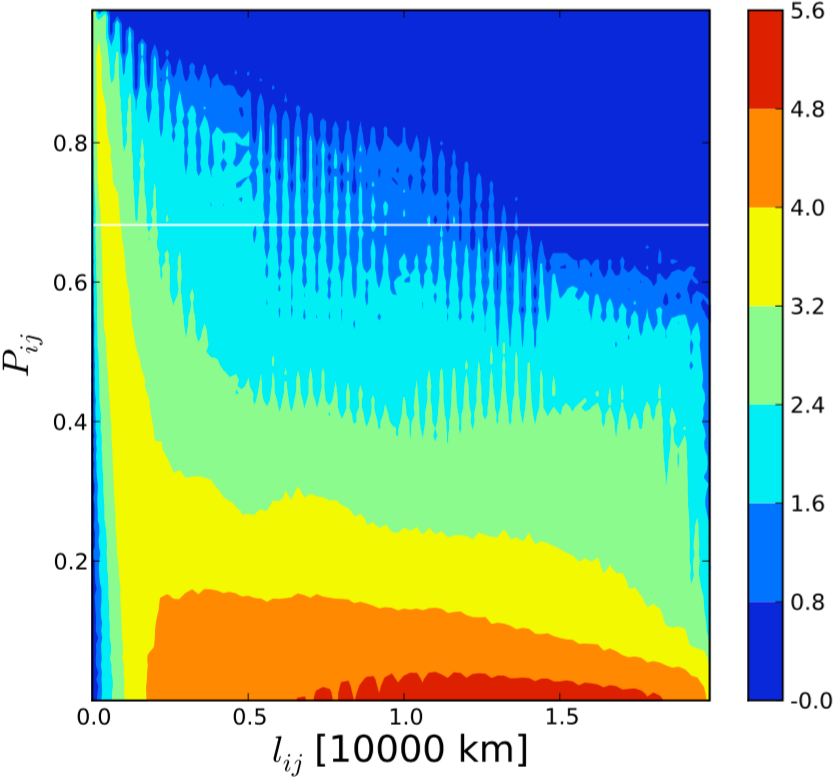}}
	\subfigure[$M_{ij}$ vs. distance]{\label{fig:clim_b}\includegraphics[width=.25\columnwidth]{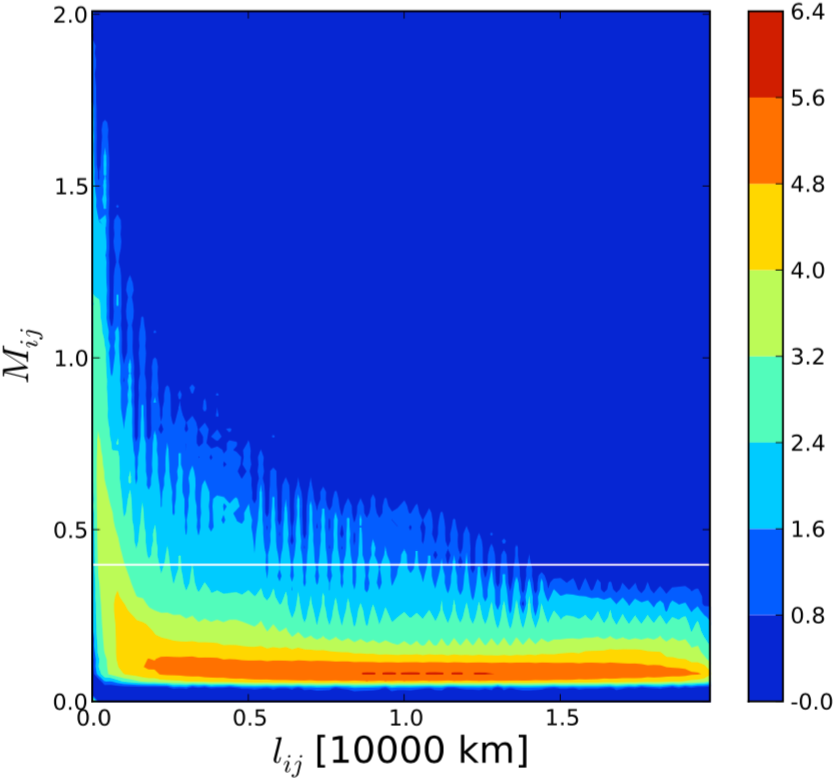}}
	\subfigure[$P_{ij}$ vs. $M_{ij}$]{\label{fig:clim_c}\includegraphics[width=.25\columnwidth]{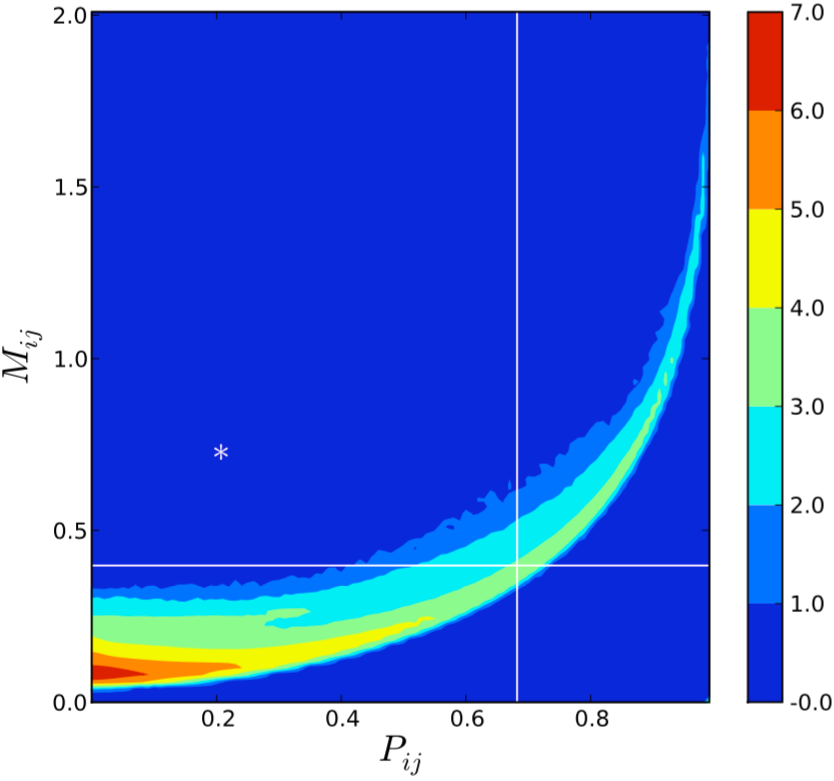}}
\caption{On the input data of averaged global surface air temperature (SAT) at different spatial regions, (a) shows the distribution of pairwise linear correlation measures ($P_{ij}$) vs. geographic distance between nodes, on a logarithmic color bar scale. This illustrates a strong spatial autocorrelation between nearby points. (b) shows the pairwise distribution for mutual information pairwise calculation vs. geographic distance, showing less spatial autocorrelation. The horizontal bars indicate thresholds on the $P_{ij}$ and $M_{ij}$ scales which produce the same network density ($\rho=0.005$). (c) shows the linear correlation vs. mutual information. The starred quadrant (top-left) denotes edges defined by mutual information but not by correlation. (From \citet{Donges2009}).}
\label{fig:climate}
\end{figure}

%
%

%

Figure \ref{fig:clim_a} shows the density of pairwise linear correlation measures ($P_{ij}$) vs. geographic distance between nodes, on a logarithmic color bar scale. This illustrates a strong spatial autocorrelation between nearby points. Figure \ref{fig:clim_b} shows the pairwise distribution for mutual information pairwise calculation vs. geographic distance, showing less spatial autocorrelation. The horizontal bars indicate thresholds on the $P_{ij}$ and $M_{ij}$ scales which produce the same network density ($\rho=0.005$). Figure \ref{fig:clim_c} shows the linear correlation vs. mutual information. The starred quadrant (top-left) denotes edges defined by mutual information but not by correlation. This demonstrates that under a fixed network density, the chosen similarity measure can yield a very different network structure.

\begin{figure}	
\centering
\subfigure[Betweenness vs. Degree]{\label{fig:climate_2a}\includegraphics[width=.25\columnwidth]{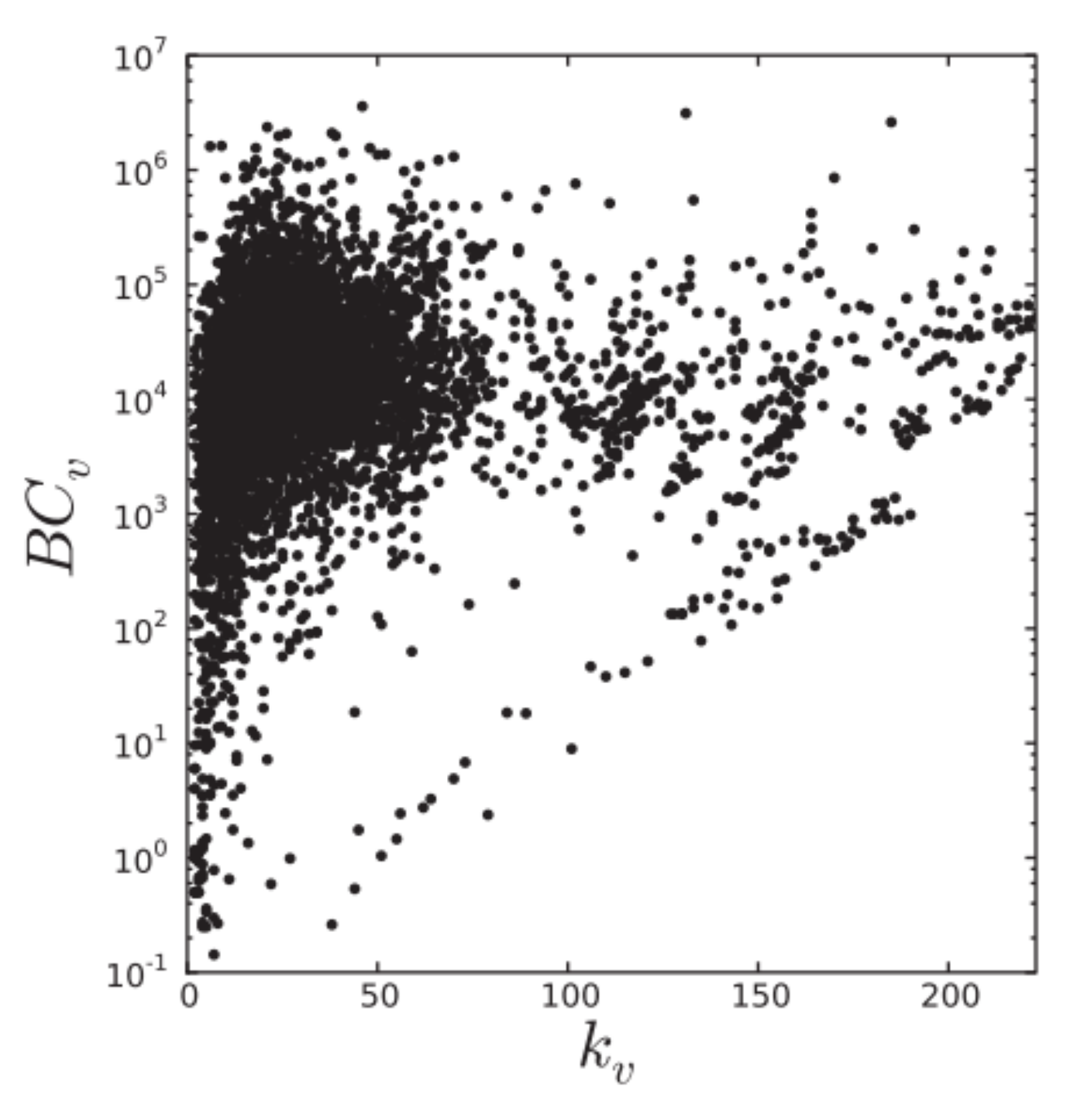}}
\subfigure[Betweenness vs. Closeness]{\label{fig:climate_2b}\includegraphics[width=.25\columnwidth]{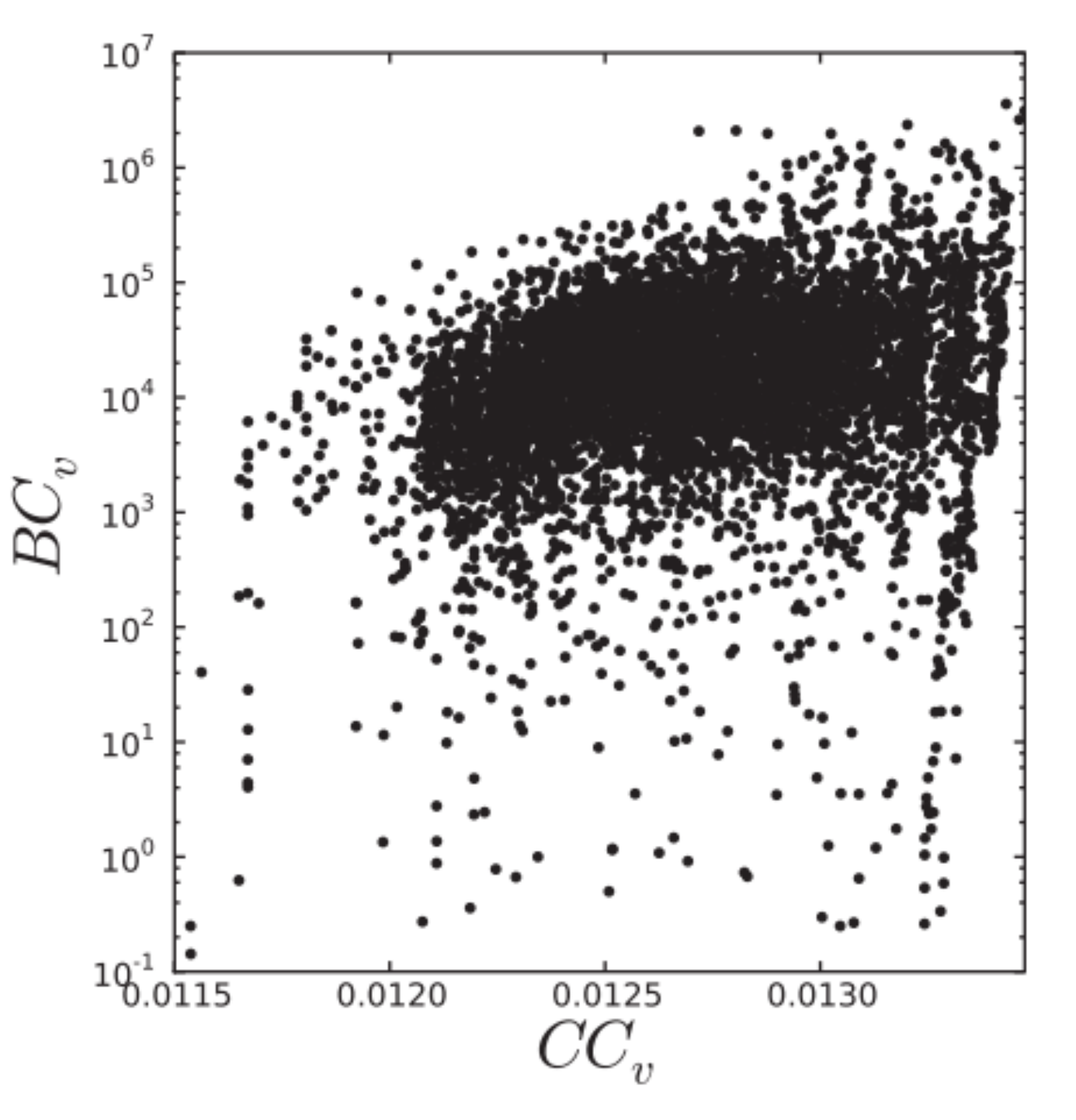}}
\caption{(a) the distribution of betweenness centrality of nodes vs. degree of nodes. (b) the distribution of betweenness centrality of nodes vs. closeness centrality of nodes. (From \citet{Donges2009}).}
\label{fig:closeness}
\end{figure}

While several methods have been developed to test edge significance, little work has focused on the validity of higher-order structures such as paths or communities. While the graph \emph{definition} of paths or communities are valid on these networks, no known work in this domain measures the \emph{interpretation} of these relationships with respect to the original time series data. 

Figure \ref{fig:closeness} explores the sub-spaces of different node measures on the inferred network. Figure  \ref{fig:climate_2a} shows the distribution of betweenness centrality per node, vs. the degree per node. Similarly, \ref{fig:climate_2b} shows betweenness centrality and closeness centrality. The authors demonstrate that degree-preserving edge re-wiring randomization indeed destroys the rank-order correlation between the marginal distributions of the node measures. 

There are two drawbacks of this analysis which re-occur across domains. First, while this methodology tests some global relational structure of the network, we are unable to interpret the relationship between any two nodes at a high geodesic distance ($\geq 2$). This means that we cannot measure properties we associate with networks, such as flow or routing. Second, significance analysis is done at a particular threshold setting, without a sensitivity analysis on the original threshold choice. In the machine learning settings, the parameter sensitivity will often be on the \emph{prediction model} parameters at a particular network definition threshold. 





\subsection{Neuroscience: Describing Functional Brain Structure and Its Connections}
\label{subsec:brain}
Biological research suggests that the brain activates interconnected, often spatially distant, regions along neuronal pathways \cite{Sporns2014}. This interconnected complexity makes networks a very natural model to study the brain. These studies are broadly in two areas: structural and functional brain networks. \emph{Structural networks} (also known as `effective connectivity', tractography, or the brain connectome) map the physical axon pathways between neurons, which may be relatively long and spatially distant. With some simplification, these networks are analogous to the physical layer in communication networks, where nodes are \emph{explicitly} connected by cables and routers. \emph{Functional networks} are analogous to the logical layer in communication networks. These networks model \emph{how} neuronal signals (\emph{i.e.} `traffic') flows over this physical layer in order to activate other neurons (\emph{i.e.} `resources') to perform different types of behavior such as auditory, visual, or motor behaviors. Researchers are only starting to understand the underlying routing and information-seeking on this physical network, and the complex contexts which change how the behavior is realized within the functional layer. Researchers aim to better understand and predict this routing, as well as the collective activation dynamics in different areas of the brain. 

Much work compares the topology of structural and functional networks using descriptive network statistics and higher-order structures, such as cluster and communities  \cite{Reijneveld2015, Rubinov2010, Sporns2016}, especially under different experimental conditions that may affect these structures, such as spinal cord injuries, epilepsy, or schizophrenia. However, all of these studies infer the network models differently, therefore it is an open challenge to rigorously synthesize these results.

\subsubsection{Underlying Data in Brain Networks}

The underlying data for structural or function networks are primarily derived from two sources. First, bio-medical imaging technologies, including Magnetic Resonance Imaging (MRI), functional MRI (fMRI), and Diffusion Tensor Imaging (DTI), detect structure of varying densities and water content. These procedures produce a flat two-dimensional image of pixels, or a three-dimensional space of voxels (often as a series of imaging over time). For example, DTI is used to construct structural networks. These images can accurately trace axon tissue connectivity by measuring flow vector orientation through the voxel space. fMRI similarly measures blood flow to voxels, a surrogate for `activity' at this location. Inferring a functional network on fMRI data then amounts to measuring statistical interactions (\emph{e.g.,} correlation) between activations in different brain areas. Second, non-invasive sensors, including electroencephalography (EEG) and magnetoencephalography (MEG), measure and localize electrical current at a probed location. Typically, these probes yield fewer and spatially coarser nodes than those defined from fMRI voxel data. These techniques produce time-series estimating electrical current (a measure for `activity') at reference locations. 

\begin{figure}
\centering{\includegraphics[width=.8\columnwidth]{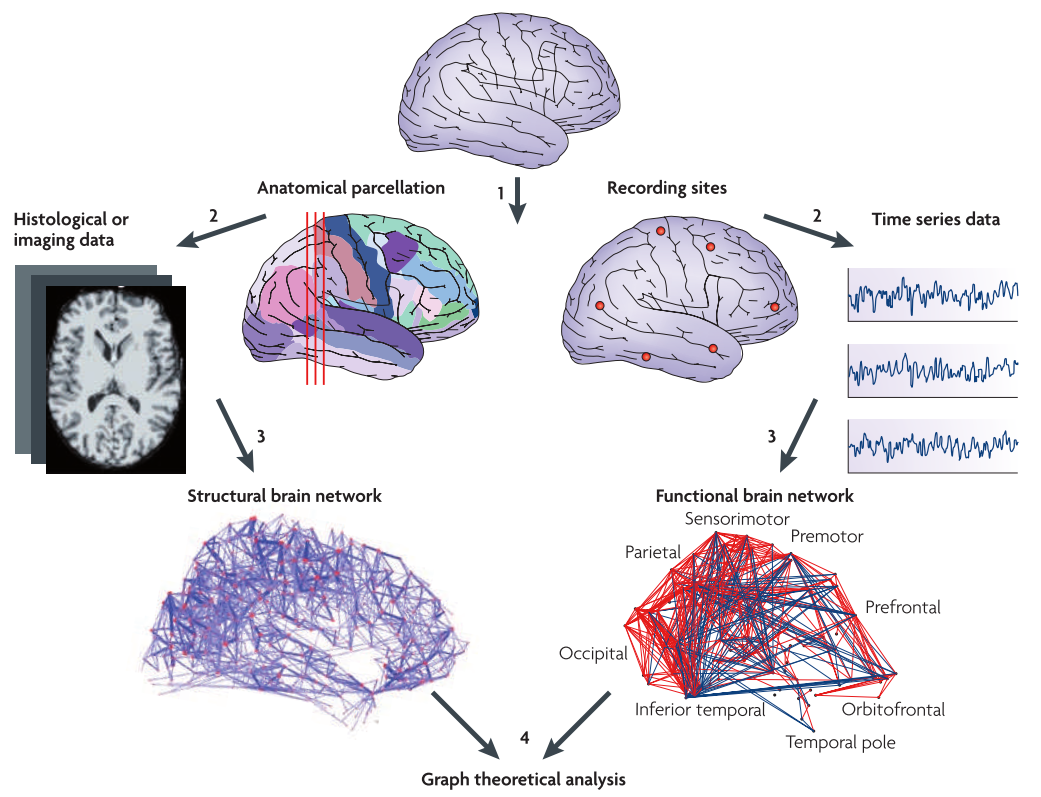}}
\caption{A methodology for constructing brain networks. (1) for a subject or population of subjects, (2) sensing techniques such as Diffusion Tensor Imaging (DTI) or functional Magnetic Resonance Imaging (fMRI) are used to sense connectivity or activity of the brain, respectively. (3) Given sensed data at recording sites (nodes), edges are inferred by different measures on the underlying data (see: Section \ref{subsubsec:brainmethods}). (4) Subsequent scientific study is conducted, using the network as a data model. (From \citet{Sporns2014}).}
\label{fig:brain_data}
\end{figure}

Figure \ref{fig:brain_data} illustrates defining both structural and functional networks, from (1) the data collection on individual subjects to (4) the final analysis task. The left path of Figure \ref{fig:brain_data} illustrates constructing structural networks. (2) \emph{Anatomical parcellation} techniques use DTI or similar imaging to determine physical connectivity in the brain. These techniques are very accurate in recovering tracts of connectivity, unambiguously. (3) these tracts are translated into nodes and edges, where previous work shows significant effects of node definition on descriptive measures such as average path length and clustering coefficient \cite{Zalesky2010}. Finally, (4) researchers use the networks as models to ask questions about the brain of the original subject or population.

The right path of Figure \ref{fig:brain_data} illustrates inferring functional networks from sensed neuronal activity. This activity can be for a range of stimulus such as music preferences \cite{Wilkins2014}, image/language associations \cite{Papalexakis:2014:GBM:2623330.2623639} or for experimental conditions such as an Alzheimer's patient cohort \cite{10.1371/journal.pcbi.1000100}. (2) fMRI, EEG, or MEG sensors measure activity at different \emph{recording sites} (contact locations, pixel or voxel locations). As in structural network construction, pixel aggregation or node definition mapping may be applied. (3) These activity response signals are compared between recording sites (nodes) with time series similarity measures (\emph{e.g.,} cross-correlation). `Sufficiently' similar time series are interpreted as latent connections between nodes, yielding the final functional network. 

One class of experiments focuses on functional networks from fMRI, coupled with a particular experimental user task (\emph{e.g.,} speaking, listening, motor). Another class focuses on inferring \emph{resting}-state networks (RSNs) of the brain \cite{Greicius07012003}. These networks are constructed in much the same way as other functional networks, except this resting connectivity is informative of very robust functional clusters. Functional networks for different user tasks can then be characterized at a higher level (\emph{e.g.,} cognitively `difficult' tasks) by comparing to the resting-state network (RSN).  

\begin{figure}
\centering{\includegraphics[width=.90\columnwidth]{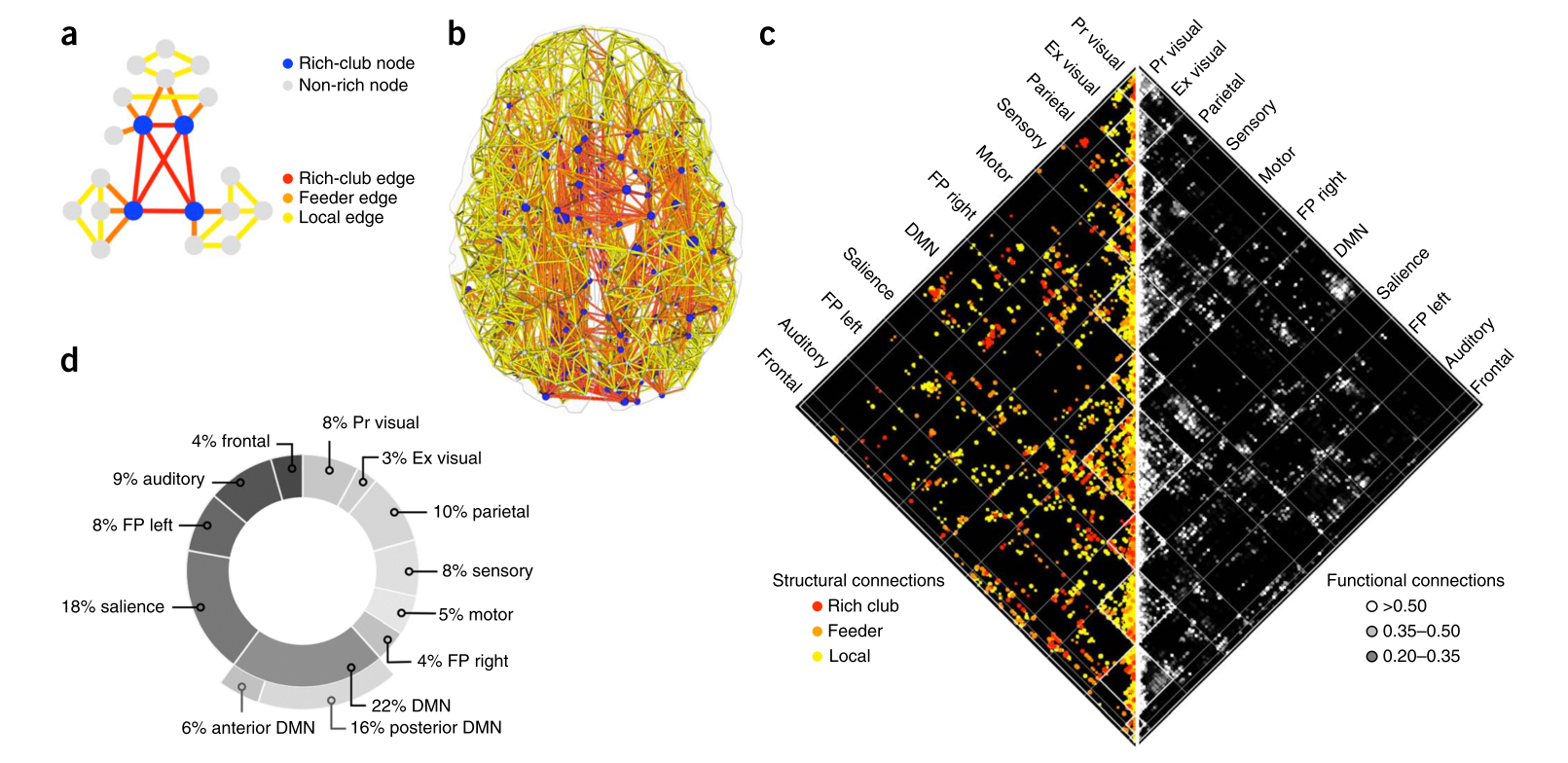}}
\caption{A study methodology comparing structural and functional brain networks. (a) a schematic of the rich-club structural network. Blue nodes indicate the $k$-core decomposition of the structural network ($k > 10$), grey nodes indicate non-rich nodes. Red (`rich-club') edges connect two rich-club nodes, orange (`feeder') edges connect exactly one rich-club and one non-rich node, yellow (`local') edges connect two non-rich nodes. (b) the network with colored nodes and edges, visualized in a brain coordinate system. (c) an adjacency matrix comparing structural topology sensed from Diffusion Tensor Imaging with rich-club edge coloring (left) to three thresholded values of functional connections for resting brain state inferred from fMRI for the same node-set (right). These nodes are ordered according to brain function in different regions of the brain. (d) the distribution of rich-club nodes within these different labeled regions. (From \citet{Sporns2014}).}
\label{fig:brain}
\end{figure}

Another network of particular research interest is the `rich-club' structural sub-networks \cite{PMID:22049421, VandenHeuvel2012, VandenHeuvel2015}. This network is essentially a $k$-core decomposition of the structural network, which indicates the global `backbone' of connectivity (where $k > 10$ is set in comparison to degree-preserving randomized networks). Nodes within the rich-club network are used to characterize the broader network into `rich-club edges' connecting two rich-club nodes, `feeder edges' connecting exactly one rich-club node, and `local edges' which connect two non rich-club nodes (see: Figure \ref{fig:brain}).  Analogous to routing in communication networks, information can flow locally within one local region for a particular behavior, or routed through backbones to physically distant regions. 

\subsubsection{Methods for Inferring Networks}
\label{subsubsec:brainmethods}
Neuroimaging time series are the dominant underlying data in neuroscience, therefore methods for constructing functional brain networks are almost exclusively in the domain of thresholded pairwise similarity measures, with some exceptions of parametric network modeling \cite{Klimm2014}. \citet{Sakkalis:2011:RAT:2072678.2072767} provides an in-depth review of these different measures, including cross-correlation \cite{Bialonski2011, Zalesky20122096}, frequency domain analysis such as discrete Fourier transform (DFT) and discrete wavelet transform (DWT) and domain-driven `coherence' measures \cite{Lachaux2002, Zhan2006, Ponten2016, PMID:10600019}. Finally, significant work focuses on causal models \cite{Ramsey20101545}, including Granger causality \cite{Roebroeck2005230,Dhamala2008354} and dynamic causal models (DCM) \cite{Rosa201266, Friston2011, David2008}.

Where these methods have threshold parameters, $\tau$, they are often validated by measuring \textit{robustness} of network statistics across varying thresholds \cite{Kramer2008173}, then using these thresholds for distinguishing patient cohorts by label or network statistic distribution. For example, previous work uses a paired $t$-test or other simple statistical test \cite{Supekar2008}. \citet{PhysRevE.79.061916} proposes a bootstrapping \cite{efron93bootstrap} method in the frequency domain, providing more general $p$-values without model assumptions. 

\subsubsection{Dynamic Functional Brain Networks}

Since the underlying data of functional brain networks is often time series, a time series of networks (dynamic networks, or time-evolving networks) are a natural extension in this domain \cite{Hutchison2013}. Network construction using time-series similarity measures (\emph{e.g.,} cross-correlation) generalize to the dynamic setting, computing on time-series subsequences. The advantage of introducing the complexity of dynamics is discovering distinct connectivity `states' over the course of the experiment. Because fMRI response can change very quickly as activity occurs over the brain, these states are lost under global time series measures \cite{Yu2015345, Robinson2015274, Damaraju2014298}. Challenges of network validation generalize to this dynamic setting, with the added challenge of appropriate \emph{temporal scale} \cite{Sulo:2010:MST:1830252.1830269, mlg2017_17}. 

\subsubsection{Comparing Functional and Structural Brain Networks}

Figure \ref{fig:brain} illustrates a complete case study summarizing many of the topics discussed above. This work integrates structural networks across 75 individuals, sensed from Diffusion Tensor Imaging (DTI) with functional networks sensed from fMRI in resting state using Pearson correlation. These different views of the network enable researchers to study how function and physical connectivity are correlated. (a) illustrates a schematic layout of \emph{rich-club} nodes, feeder, and local nodes. (b) shows the layout of these nodes in a brain coordinate system, with the same node and edge coloring. (c) illustrates an adjacency matrix comparing structural connections (left) with three thresholded values of functional connections within the same node-set (right). Furthermore, nodes in different spatial regions of the brain (\emph{e.g.,} default mode network, motor, auditory, frontal) are labeled according to primary function, showing structural and functional edges between these regions. (d) shows the distribution of rich-club nodes within these different labeled regions.

\subsection{Epidemiology, Blogs, Information Networks: Modeling Virus Spread and Information Flow}
\label{subsec:information}
\begin{figure}	
\centering
\subfigure[]{\label{fig:epi1}\includegraphics[width=.17\columnwidth]{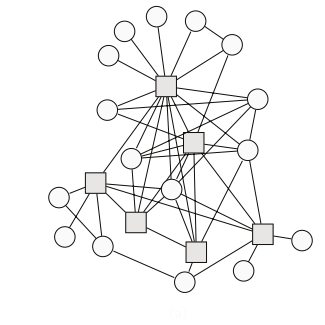}}
\subfigure[]{\label{fig:epi2}\includegraphics[width=.17\columnwidth]{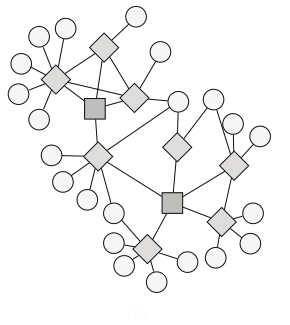}}
\subfigure[]{\label{fig:epi3}\includegraphics[width=.17\columnwidth]{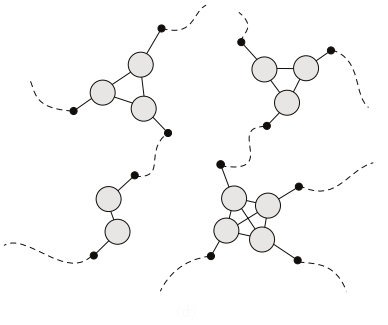}}
\subfigure[]{\label{fig:epi4}\includegraphics[width=.17\columnwidth]{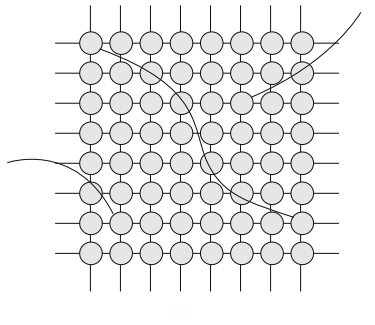}}
\caption{Examples of networks inferred from data, or modeled in the epidemiology domain. (a) depicts a contact network where squares are respondents and edges are sexual or drug contact which might transmit HIV \cite{Bell19991}. (b) a sexual network derived from snowball sampling of respondents (squares), where the edges between non-respondents (circle nodes) are unknown. (c) a network of households (cliques) and their interconnections for modeling infection in realistic social contact networks. (d) The `small-world' network property modeled through a lattice with sparse edges connecting distant nodes. (From \citet{Danon2011}).}
\label{fig:epidemology}
\end{figure}

\subsubsection{Networks in Epidemiology}

Networks are used in epidemiology to simulate the spread of disease over a family of parametric network models (\emph{e.g.,} random, small-world, exponential random graphs) representing contact between entities over time \cite{Keeling2005}. Network structure inference in epidemiology and information networks aims to discover an unobservable network (\emph{e.g.,} physical contact networks, sexual networks, malware transmission) over which information or infection is spread. We observe the \emph{effects} of transmission on the infected node (e.g symptoms), but not the edges over which they spread. The focus in this domain is therefore parametric inference for these models from data.

Modeling \emph{contact networks} allows researchers to simulate different outbreaks on these networks. Figure \ref{fig:epidemology} illustrates types of network data and models used in the epidemiology domain \cite{Danon2011}. Figure \ref{fig:epi1} visualizes a contact network from survey data, where squares are respondents and edges identify the transmission risk for HIV through contact by drug use or sex \cite{Bell19991}. Figure \ref{fig:epi2} shows a snowball sample of respondents (as squares) and their partners. In this example, edges between circle nodes are unknown. Subsequent modeling for edges between circle nodes can test the spread over this population under different unobserved contact assumptions. Figure \ref{fig:epi3} illustrates a model of households (cliques), sparsely connected to others. This is intended to model contact networks and potential spread through family-unit environments. Finally, \ref{fig:epi4} illustrates a lattice network with sparse edges outside of the local neighborhood. This model was previously used to capture the `small-world' property of information and disease spread \cite{Boots1933}.

Historically, inferring networks in this area focuses largely on parametric graphical modeling using Markov Chain Monte Carlo (MCMC) \cite{brooks2011handbook} and maximum likelihood methods, incorporating modeling assumptions in transmission and interaction rates. 

Given a compartmental transmission model \cite{Brauer2008} (\emph{e.g.,} the susceptible-infected-recovered SIR model or susceptible-exposed-infected-recovered SEIR model), these methods measure the likelihood of possible \emph{sequences}, or trees of infection, where infection times from a root are monotonically increasing. Let $t_i$ and $t_j$ denote the infection times of nodes $n_i$ and $n_j$, then the transmission model yields the likelihood $P(\text{``j infected i''} | t_j - t_j)$. 

\subsubsection{Contact Network Inference}

Early work in the epidemiology domain focused on the inference of either spread parameters (\emph{e.g.,} infection rate) or network model parameters on random graphs \cite{BRITTON2002}, Poisson, and power-law networks \cite{Meyers200571}, as well as fitting of real-world data to a contact network model \cite{Bansal879}. In these analytical and simulation results, the network model is known and no structural inference is necessary. These studies generally model the spread of an epidemic under possible individuals in contact (called `contact tracing') \cite{Patrick01042002} and discovering the root individual(s) of the infection (called `transmission tracing'). Early work also formulates association network heuristics based on time and distance of potential contacts \cite{Haydon2003}.

\subsubsection{Infection-Time Cascades}

Previous work in machine learning uses network structure inference to represent the spread of information between nodes, where the edges of transmission are unobservable. Maximum-likelihood methods have focused on learning a network under assumed transmission rate models, using statistical inference for these parameters \cite{NIPS2010_4113}.

\begin{figure}
\centering
\subfigure[]{\label{fig:cascade1}\includegraphics[width=\figtwoscale\columnwidth]{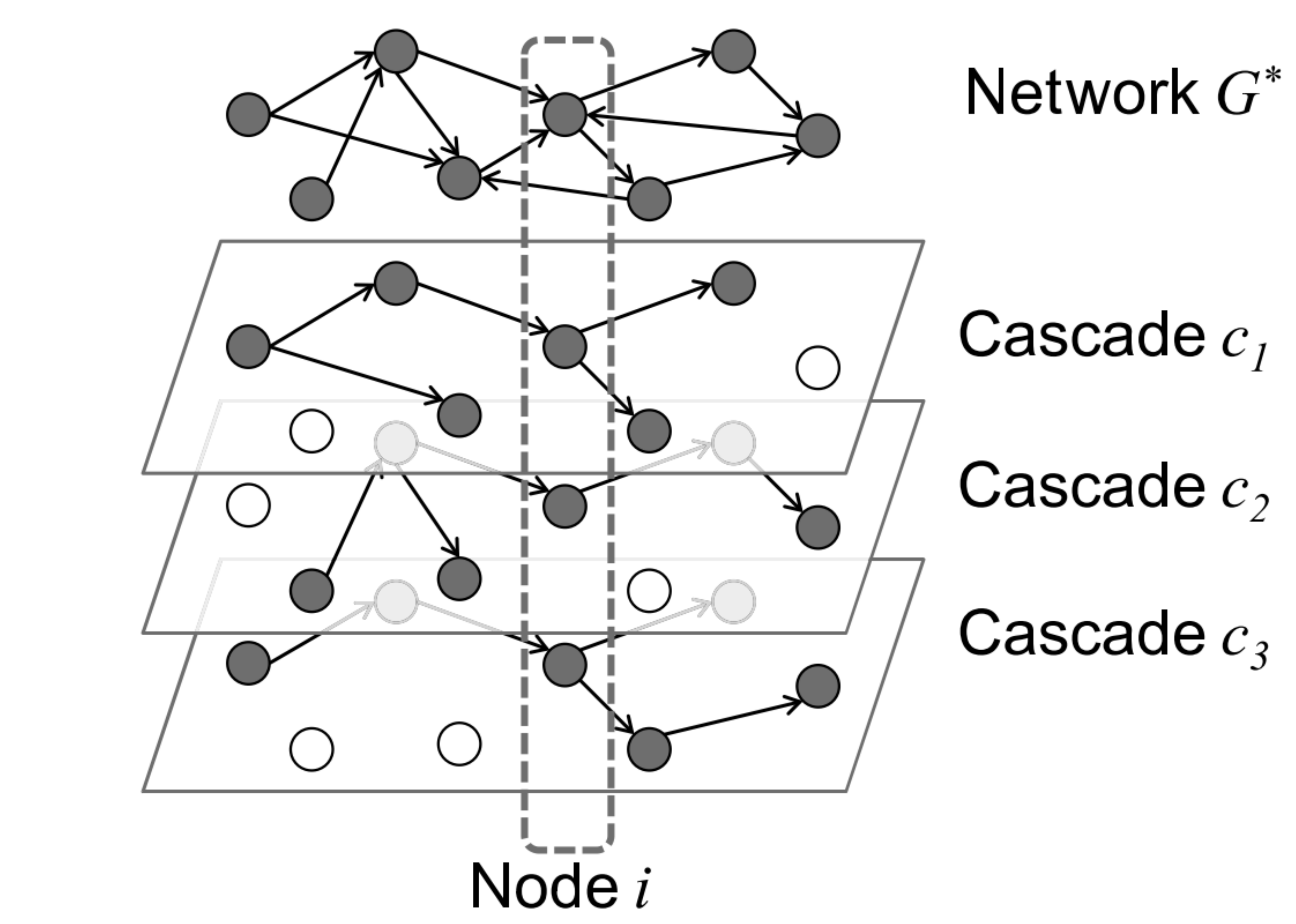}}
\subfigure[]{\label{fig:cascade2}\includegraphics[width=\figtwoscale\columnwidth]{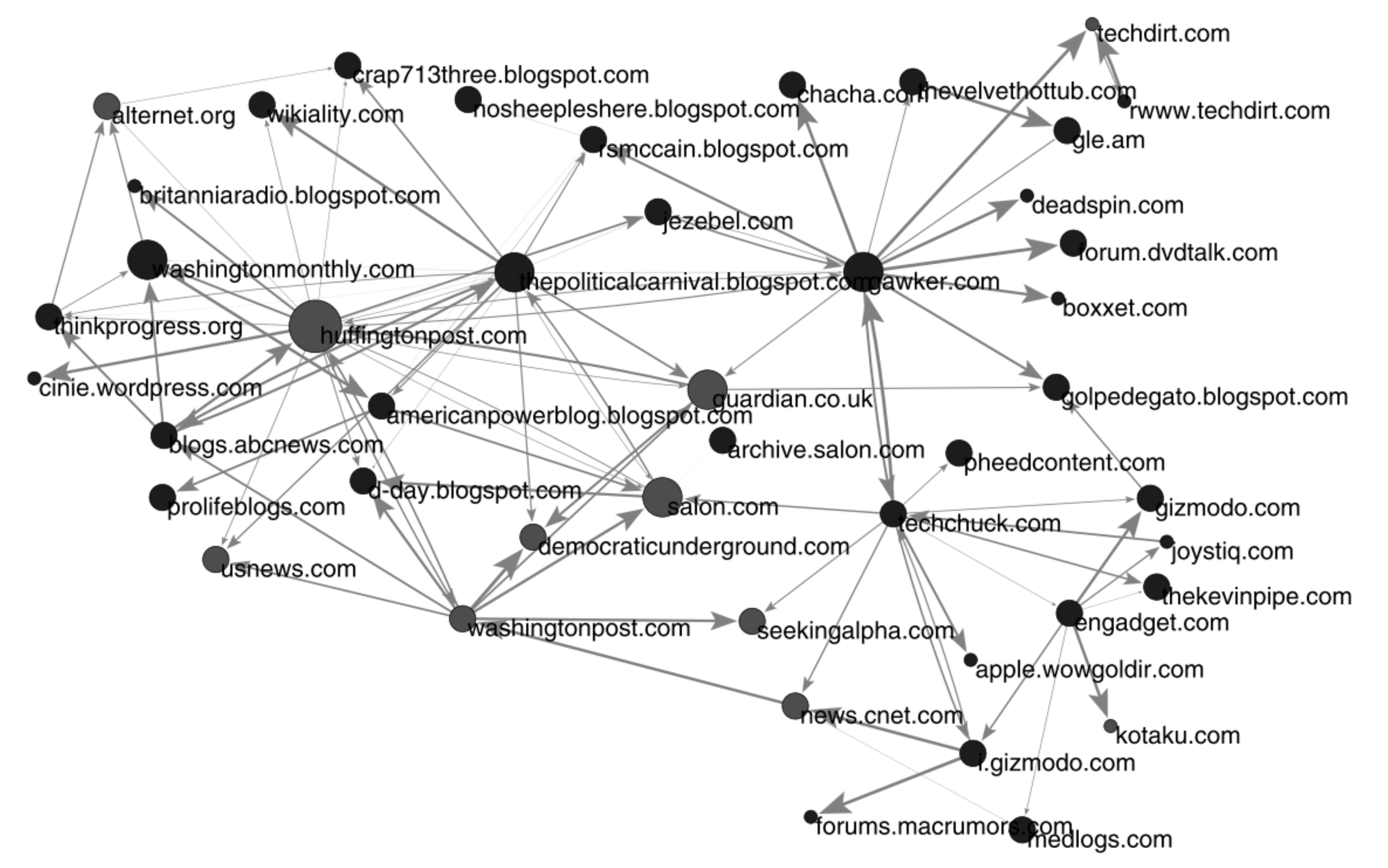}}	
\caption{(a) illustrates constructing network $G^*$ on a collection of cascades $c_1, c_2... c_k$. For a source node $s$, these methods estimate the likelihood of infecting $i$ given the observation of information arrival (`infection') at node $i$. (b) An estimated network inferred from hyperlink arrival times at nodes. (From \citet{Gomez-Rodriguez:2012:IND:2086737.2086741}).}
\label{fig:diffusion_ex}
\end{figure}

Figure \ref{fig:cascade1} illustrates the intuition of network construction by information propagation through unobservable edges. To recover the unobservable true network $G^*$, each sequence of non-decreasing infection times (\emph{e.g.,} ``cascades'') supports the possible transmission between nodes with sufficiently close infection times. The key step of this work is learning the parameters of a transmission model which measures the likelihood of a cascade according to differences in ``adjacent'' infection times in the network.  

\citet{NIPS2010_4113} use convex programming to learn a maximum-likelihood network under a fixed transmission time probability distributions $w(t)$ and recovery-time distribution $r(t)$. To learn the transmission weight matrix $A$, the authors use an Independent Cascade model \cite{Kempe:2003:MSI:956750.956769} where an uninfected node $n_i$ is exposed to infection by adjacent infected nodes $n_j$ at each time step using a Bernoulli process with a probability $A_{j,i}$. The authors present a convex optimization formulation of their likelihood function, with regularization. This model, and most of the subsequent work, is evaluated on synthetic network models where the underlying network is explicitly known. The `task' is the accurate reconstruction of the network which simulated these infection times. Similar to gene regulatory networks, the final evaluation of the network is the network itself, rather than a subsequent task on the inferred network. These methods then also typically present qualitative results on real-world datasets.  

\citet{Gomez-Rodriguez:2012:IND:2086737.2086741} scales this previous work by fixing a global edge transmission probability $\beta$ and solving this simplified problem. For many applications, this fixed transmission probability assumption can be made. The primary insight under this assumption is that we can simply use the most likely propagation tree over a set of nodes in a cascade $c$ (see: Figure \ref{fig:diffusion_ex}). Given a cascade set $C$, the authors marginalize their likelihood function relative to edge selection and prove this function is monotonic and submodular. Therefore, edges can be greedily selected with an approximation factor of $(1 - 1/e)$ \cite{Nemhauser1978}. 

\begin{figure}
	\centering
	\subfigure{\label{fig:pair1}\includegraphics[width=.22\columnwidth]{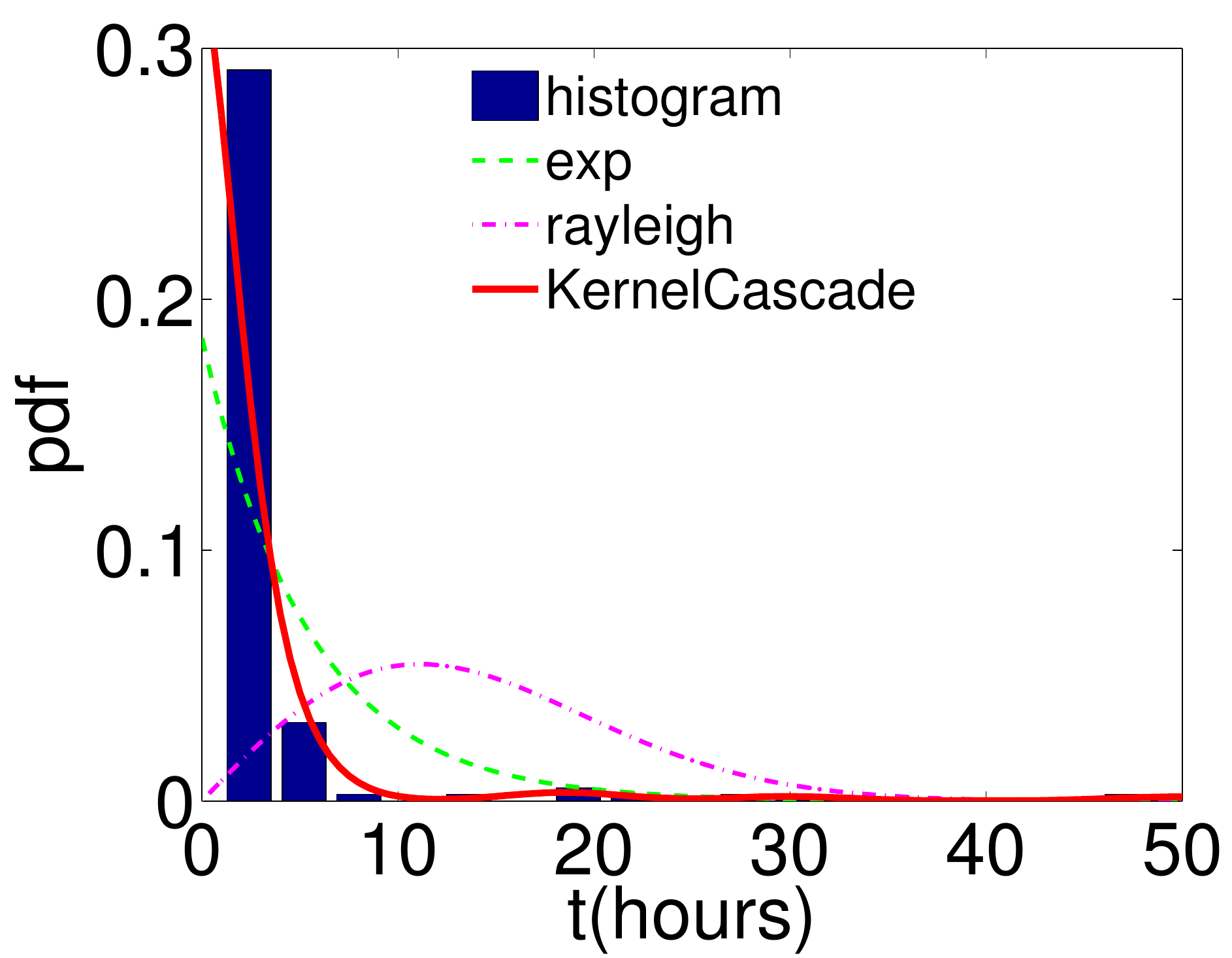}}
	\subfigure{\label{fig:pair2}\includegraphics[width=.22\columnwidth]{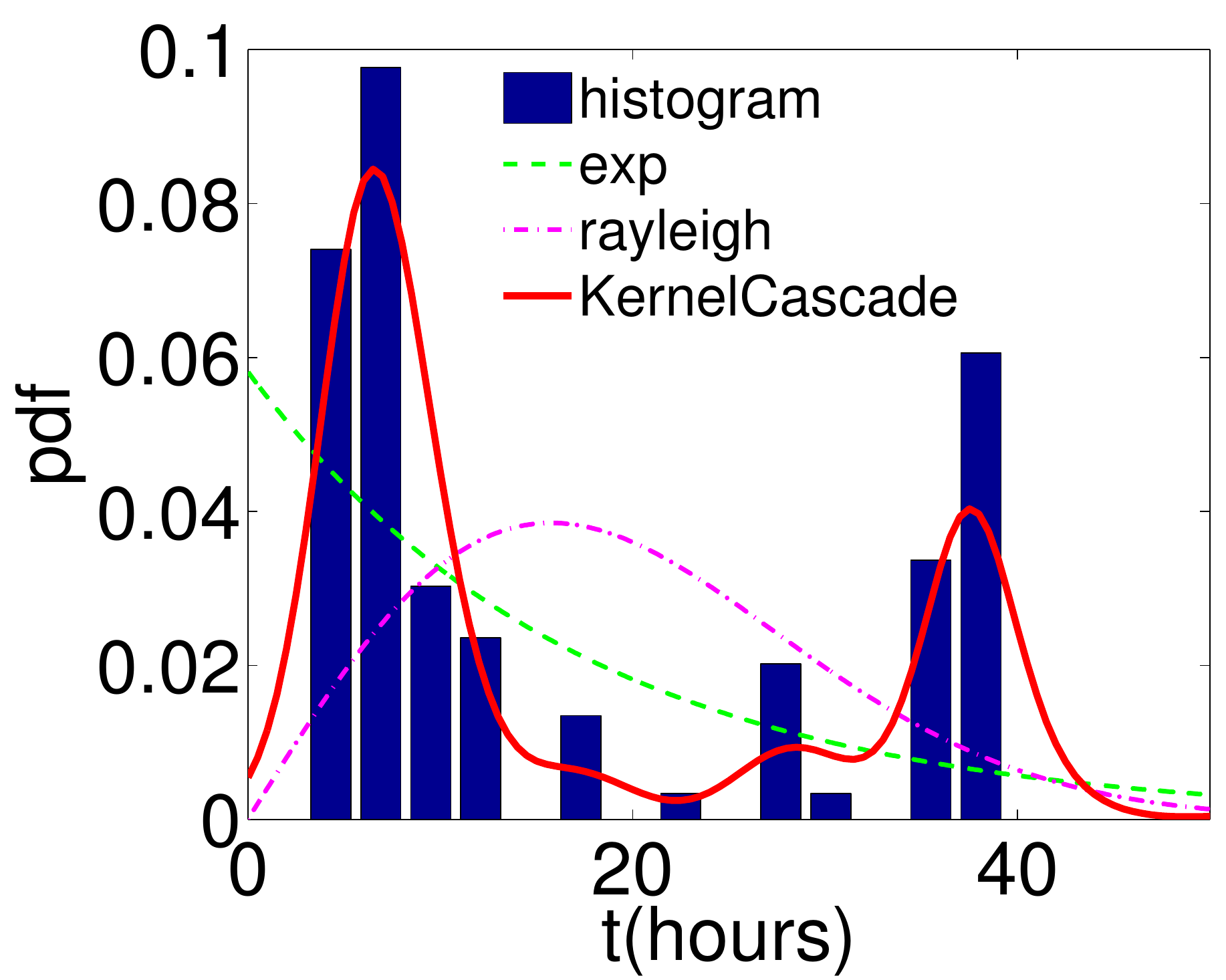}}
	\subfigure{\label{fig:pair3}\includegraphics[width=.22\columnwidth]{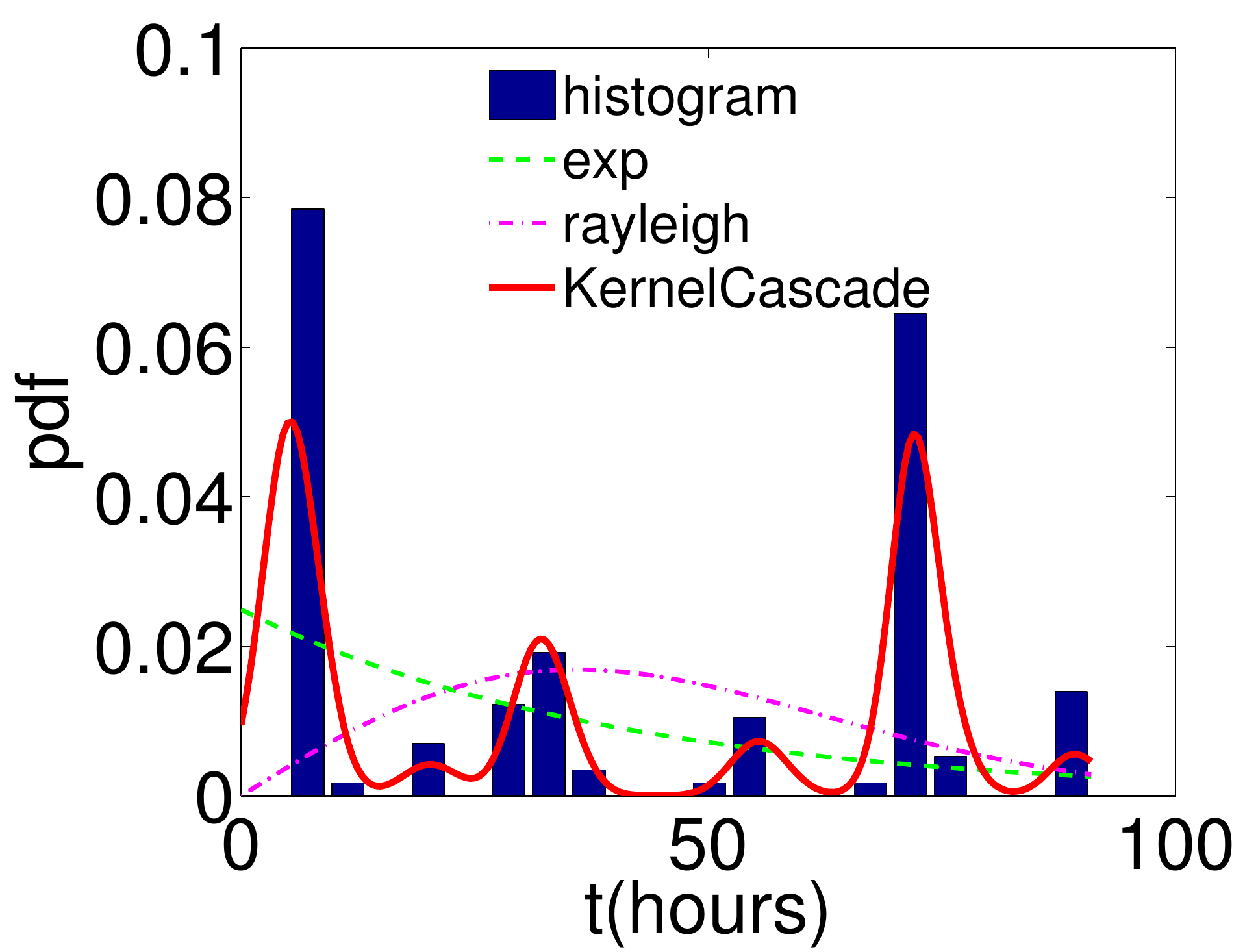}}
\caption{Transmission histograms, showing poor fit of exponential and Raleigh models. (From \citet{NIPS2012_4582}).}
\label{fig:diffusion_fit}
\end{figure}

The transmission model of \cite{NIPS2010_4113} assumes repeated Bernoulli trials of fixed probability. \citet{ICML2011Gomez_354} introduce transmission likelihood functions which vary over time. Given a set of cascades $C$, the method infers the parameters of transmission rate models for each edge. The model uses a hazard function $H(\tau_i|\tau_j; \alpha_{j, i})$ that measures the instantaneous infection rate on node $n_i$ from infected node $n_j$, given the parametric function on $\alpha$. The authors present three different transmission model functions: Exponential, Power-law, and Rayleigh, and prove that the optimization for transmission rates matrix $A$ is convex. 

In many real-world applications, information can propagate in a multitude of ways. Older information might be promoted by an influential node, causing multi-modal spikes in propagation after some delays. Therefore, these transmission rates may not decay monotonically, nor according to any simple parametric function. 

Figure \ref{fig:diffusion_fit} illustrates this intuition on real transmission data between blog sites. The figure reports the probability density function ($y$-axis) over varying time delays between post creations on different nodes ($x$-axis). The data poorly fits any single exponential or Rayleigh transmission model. \citet{NIPS2012_4582} extends previous work to address this limitation. The key addition of this work is to kernelize the hazard function $H(\tau_*|\tau_j, \alpha_{j,i})$ over $m$ different kernels. These kernels serve as a piecewise approximation of the time-lag distribution, which can then be used to estimate the likelihood of transmission between nodes, given observed `infection' data. 

We have dug deeper into the technical details in this section to demonstrate the added modeling complexity accounting for richer assumptions of the underlying transmission process. However, there is still an important gap. Previous work does not typically account for null modeling. In this instance the underlying process may not be a network, but a confounding process which might individually `infect' nodes.

\subsection{Ecology: Inferring Animal Social Networks to Explain Social Behavior}
\label{subsec:ecology}
\subsubsection{Networks in Ecology}

Networks in ecology model systems at differing levels of abstraction. In systems ecology, traditional graphical models are used to model an ecosystem at a high level, with relationships between species, environmental variables, services, and other processes  \cite{Milns2010}. For example, food webs model who-eats-whom within an ecosystem \cite{Proulx2005345}. A second application of networks in behavioral ecology model social systems of animal populations \cite{Wey2008333,JANE:JANE12418}. These networks are the focus of this section.

Analogous to sociology and political science, traditional fieldwork data in ecology are collected from direct observation and `surveys,' measuring interactions or other relationships among individuals in the population \cite{Sueur2011, Barrett2108}, particularly over time \cite{Pinter-Wollman14062013, DeSilva2011,McDonald2007}. Networks serves as models to summarize and describe these social or behavioral relationships within these populations.

Direct observation of populations allows researchers to incorporate their own intuition and experience into the definition of these networks. In practice, much of the work in this area uses ad-hoc, intuitive network definitions with some sensitivity analysis. These networks derived from direct observation are typically categorical (\emph{e.g.,} kinship relations) or discrete (\emph{e.g.,} thresholding on number of interactions, where interactions are implicitly decided by observers). Methodological consideration are well established in the field on these data, including edge strength thresholding \cite{Croft2009}, sampling, hypothesis testing \cite{Croft2011502}, and randomization strategies \cite{James2009, Haddadi2011}. Each of these provide several different choices for validating the definition of these networks. These networks are tightly coupled with a particular hypothesis and experimental cohorts in the population. 

Two networks of interest in animal social networks measure \emph{affiliations} and \emph{associations} between individuals \cite{MEE3:MEE312383}. Affiliations describe \emph{intentional} social relationships between individuals \cite{Croft2011502, Whitehead2005e1} (\emph{e.g.,} grooming pairs of primates), while associations describe a broader set of interactions, which might be driven by structural factors rather than social affinity (e.g. environmental resources, sex, age, and other individual attributes) \cite{BEJDER1998719}. \citet{MEE3:MEE312383} introduce a generalized affiliation index using a linear regression model, using it to remove sets of predictive structural features. The resulting model by subtraction is the affiliative network model.

\subsubsection{Instrumentation and Sensing of Animal Populations}

Recent instrumentations of individuals and the environment allow the observation of ecosystems and populations at an unprecedented scale using geo-location sensors such as GPS, proximity sensors, radio-telemetry, and Passive Integrated Transponder (PIT) tags \cite{Krause2013, Kaysaaa2478, Rutz2012R669}, as well as individual identification by photographs \cite{2017arXiv171008880B}. This instrumentation allows the study of detailed individual behavior and social dynamics that are outside the direct observation of researchers. This abundance of data requires novel statistical techniques for inferring networks from \emph{implicit} interactions. 


Defining meaningful `interactions' directly from data is non-trivial. No known work compares the biases of interaction and/or affiliation sampling via traditional direct observation fieldwork, against the interaction or affiliation inference from underlying data. Presently, these sensors are most effective at recording simple co-location or trajectories. Challenging independent problems such as activity recognition (\emph{e.g.,} grooming, conflict) are much more easily identified by direct observation by researchers than by data, but these direct observations do not scale. Future research will likely integrate the strengths of these modalities to improve fieldwork data collection.

\subsubsection{Studies and Network Inference Methods on Instrumented Data}

The key difference between data from traditional fieldwork and from instrumented technologies is that the former tend to be discrete counts (\emph{e.g.,} number of co-locations or grooming events), while the latter are continuous data without these higher-level labels (\emph{e.g.,} relative distances between individuals). To translate to discrete co-location events--and subsequently a network--requires defining ``how close'' for ``which duration'' constitutes a co-location edge, or ``how correlated'' for ``which duration'' constitutes a higher-order edge in the network.

The simplest method for setting these closeness and persistence thresholds for co-location is by domain knowledge, or by sampling the parameter-space in some way. \citet{Haddadi2011} use this strategy in GPS data from sheep, ranging from individuals co-locating for 1 minute at 1 meter, to 5 minutes at 3.5 meters. The authors have some known `affiliations' (as described above), which are used to validate network accuracy at these different thresholds when the individuals are mixed into a larger population. \citet{Aplinrspb20121591} collected data from passive integrated transponder (PIT) tags of individuals sensed by radio-frequency identification (RFID) antennae at feeder sites.  This work defines associations as two individuals co-occurring at the site within 30 seconds before or after the other on a sliding 75 second window. Co-occurrence is categorical due to the physical design of feeders, so only the `persistence' of interaction need be defined. This threshold generates a stream of pairwise associations which is then thresholded again ($\tau \geq 0.02$) to produce an aggregated association network.

\citet{Psorakis07112012} define edges using Gaussian mixture models (GMM) on co-occurrence data for a similar feeder system. This approach mitigates the `persistence' threshold by fitting Gaussian distributions to a one-dimensional space of occurrence counts (and generalized to continuous two-dimensional geographic space, \cite{Farine2016}). These distributions capture \emph{events} of co-occurrence among several individuals. \citet{Hamede2009} used a randomization approach to define non-random associations on proximity sensors on a population of wild Tasmanian devils (\emph{Sarcophilus harrisii}). Internal thresholds on these sensors detect co-location within 30 centimeters. This work studies disease transmission through physical contact of the animals, so this thresholding is appropriate. 

We enumerate several instances to demonstrate that these network constructions are typically `ad-hoc' from domain knowledge. In each case, the sensitivity of downstream hypotheses is not tested against the choice of network threshold parameters. This may result in inadvertent fine-tuning of parameters and incorrect conclusions.

\subsubsection{Translating Co-Location Data to Innovation Spread in Networks}

\begin{figure}	
\centering
\subfigure[]{\label{fig:birds1}\includegraphics[width=.22\columnwidth]{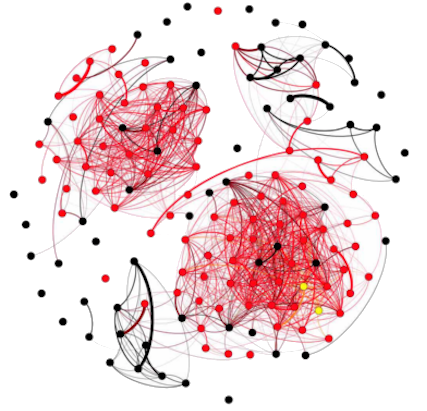}}
\subfigure[]{\label{fig:birds2}\includegraphics[width=.22\columnwidth]{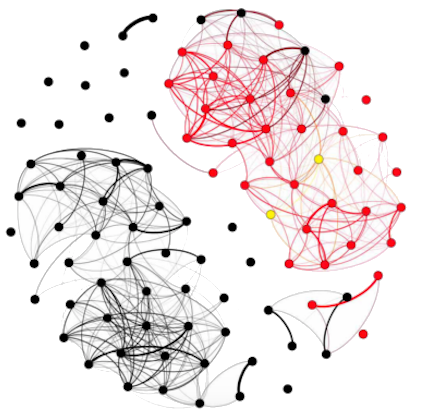}}
\subfigure[]{\label{fig:birds3}\includegraphics[width=.22\columnwidth]{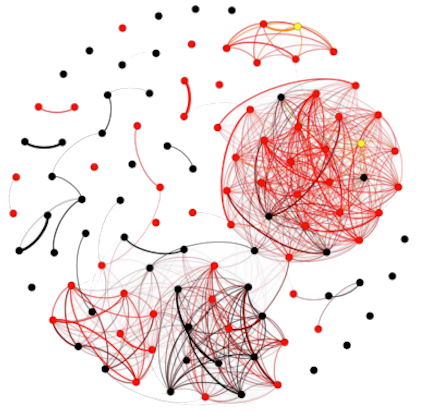}}
\caption{Networks inferred from co-location at feeding stations sensed using RFID over three different populations. These edges are colored showing learned behavior (red edges) of obtaining food through an instrumented puzzle mechanism, the trained individuals (yellow nodes), and affiliated individuals using the default strategy (black edges). (From \citet{Aplin2015}).}
\label{fig:birds}
\end{figure}

\citet{Aplin2015} proposed a network inference task to measure the learning of a feeding behavior in a population of great tits (\emph{Parus major}). In this experiment, feeders were instrumented with RFID antennae and recorded the visitation of each unique bird using PIT tags. The feeders had a sliding door that needed to be open to the left or right to access the food. The feeder recorded the bird's puzzle solution. The authors investigated whether birds learn by example from others at the feeding sites. Figure \ref{fig:birds} shows a thresholded, aggregated network over individuals, weighted by the frequency of co-location events at any feeder, using the Gaussian mixture model (GMM) method described above for interaction `events'. Yellow nodes represent trained individuals, red nodes represent individuals that have learned the correct behavior to solve the feeder by the end of the study. 

These networks visually show strong network modules between red and black individuals. Figure \ref{fig:birds2} shows a strong network separation between the two behaviors, where trained individuals are within the red cluster. Figure \ref{fig:birds3} shows two strong red clusters around both trained individuals, but also that the correct behavior has spread across a component of untrained individuals.

\subsubsection{Testing Biological Hypotheses Using Networks as Models}

\citet{Farine2016} evaluates networks inferred from geo-location data of individual olive baboons (\textit{Papio anubis}) within a troop. The authors evaluate varying networks for their ability to be used as models for predicting the future location of individuals. 

\begin{figure}	
\centering
\includegraphics[width=.35\columnwidth]{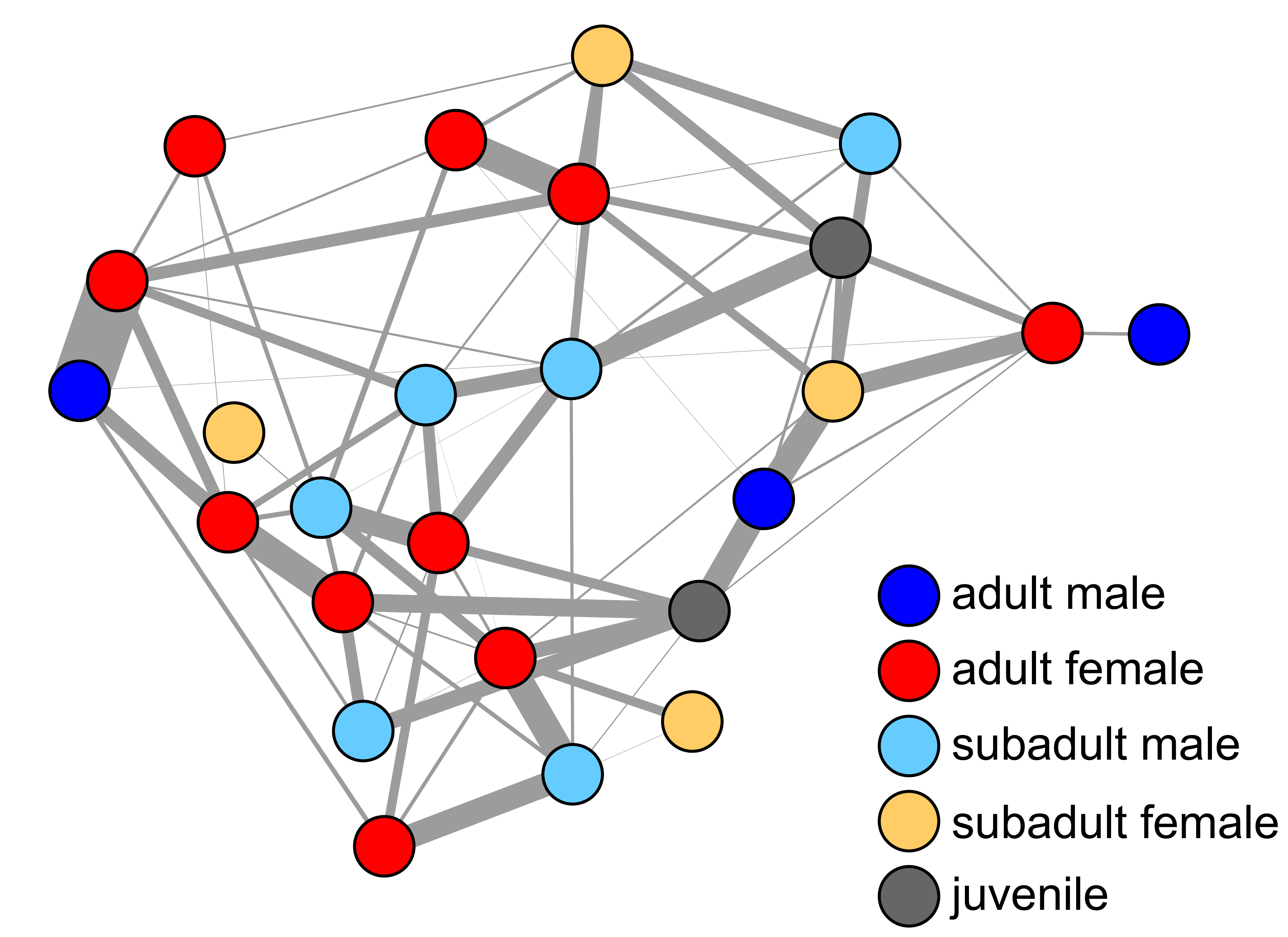}
\caption{A ``co-sitting'' network inferred from geo-location data on a population of baboons. Edge weights are proportional to the time two individuals are within 1.5 meters for at least 1 minute. (From \citet{Farine2016}).}
\label{fig:baboon_cositting}
\end{figure}

Figure \ref{fig:baboon_cositting} shows an affiliation network inferred on a large time-scale. Individual baboons are fitted with high-resolution GPS collars and are represented as nodes. Edge weights indicate the proportion of time two individuals are within 1.5 meters for at least 1 minute, over much of the course of data collection. The authors also generate an alternative $k$-nearest neighbor network model over varying $k$ at shorter time-scales. For a given node, the location centroid of \textit{neighbors} in each of these networks is used as a prediction of the individual's location.

This collection of networks represents different \emph{hypotheses} for the mechanism of predictable interactions between baboons: how many individuals do baboons pay attention to at a time? How long do these relationships persist? And do individuals have fixed preferences which are more predictive than interactions over time?

This study finds that a $k$-nearest neighbor network of $k=4,5,6$ are most predictive and statistically significant. This is biologically meaningful and agrees with previous work showing grooming cliques in baboons tend to be size 4--8 \cite{Dunbar1992}. This size indicates the biological capability of baboons to manage attention and social relationships at a given time. The study also finds that at longer time scales on the order of 10 minutes, the network of long-term affiliations becomes more predictive. So, different networks are more appropriate models of behavior on different time scales.


\subsection{Mobile Social Networks: Studying Human Mobility Through Social Relationships}
\label{subsec:mobility}
\subsubsection{Underlying Data in Mobile Networks}
Phones and other mobile devices are among the most versatile and informative sensors of personal and social activity \cite{5560598}. The modeling of mobile phone data as networks is motivated by the complex, overlapping, and dynamic modalities which are sensed by these devices. Mobile devices collect physical proximity (bluetooth, WiFi), physical location (GPS), direct communication (SMS, voice), interactions with applications, interactions through other online social networks and email \cite{klimt2004enron}. Integrating these modalities promises to give a rich picture of large-scale human mobility, dynamics and scale \cite{Saramaki2015}, geography and communication \cite{Ratti2010, blondel2010regions}, and offline face-to-face social networks. 

While much of this underlying data is similar to data collected for animal social networks (proximity, location, discrete interactions), there are notable trade-offs between privacy and experimental design within these domains. While animals are not due rights to data privacy, they are also unable to comply with instructions or be surveyed for ground truth network edges. 

In human experiments, contact diaries \cite{Mastrandrea2015} or Facebook friends \cite{Sekara2014} have been collected to validate networks inferred from proximity sensors. Experiments on mobile users are necessarily less invasive, while topics such as disease spread and sexual contact networks are often sensed in animal populations. Data privacy requires careful, informed consent and secure storage \cite{2014arXiv1403.5299S}; location has been shown very effective to uniquely identify users using only a few data points \cite{DeMontjoye2013}. 

Several mobile datasets have been collected for the purposes of social research \cite{Blondel2015}. The ``Reality Mining'' dataset is the first large-scale collection, on 100 participants (faculty and students) in the MIT Media Laboratory \cite{Eagle:2006:RMS:1122739.1122745}. This anonymized dataset contains call logs, Bluetooth device proximity, cell tower ID (a proxy for location), and other fields. Similar mobile data collection projects followed, including the Lausanne Data Collection Campaign on 170 student participants \cite{Laurila2013}, the Social fMRI study on 130 participants \cite{Aharony:2011:SFI:2072697.2073099}, and the SensibleDTU project of $1$,$000$ participants \cite{Stopczynski2014}. These subsequent studies collected more detailed user activity, surveys, online social network activity, and detailed user demographics, addressing the limitations of previous efforts. 

Finally, the SocioPatterns platform \cite{10.1371/journal.pone.0011596, 2008arXiv0811.4170B} uses a specialized proximity sensor design to record face-to-face interactions. These sensors have been deployed in an academic conference setting \cite{Smieszek2016}, elementary schools \cite{Stehlé2013604}, high schools \cite{Mastrandrea2015}, and several other environments. The specificity of these sensors for detecting individual \emph{interactions} between users addresses the challenges of using general proximity sensors (\emph{e.g.,} Bluetooth) for population studies.


\subsubsection{Studies and Methods on Mobile Data}

The primary task in inferring networks from mobile data is related to comparison across modalities, for edge or attribute prediction. Previous work focuses on predicting Facebook friends from Bluetooth co-location \cite{Sekara2014}, as well as survey-reported friends from proximity and call record data \cite{Eagle2009}.

\begin{figure}
	\centering
	\subfigure[]{\label{fig:realitya}\includegraphics[width=.4\columnwidth]{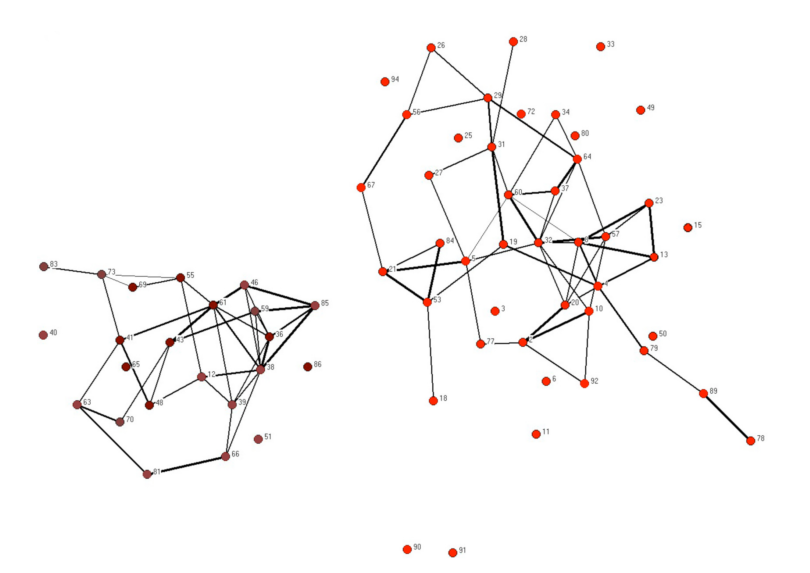}}
	\subfigure[]{\label{fig:realityb}\includegraphics[width=.4\columnwidth]{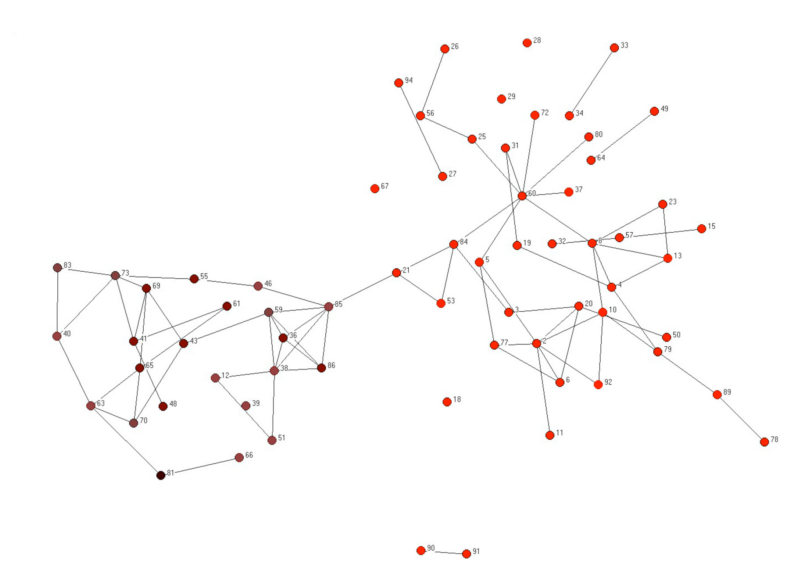}}
\caption{(a) a weighted latent network inferred by Bluetooth proximity and call records on the population of Reality Mining users. (b) the self-reported ground-truth friendship network. (From \citet{Eagle2009}).}
\label{fig:eagle}
\end{figure}

Figure \ref{fig:eagle} examines this latter task. The network in Figure \ref{fig:realitya} is inferred using principle components analysis of Bluetooth proximity counts across different times and locations (\emph{e.g.,} work, off-campus, weekday, weekend). The authors assign edge weights according to the PCA factor that by inspection corresponded to ``non-work'' hours (\emph{e.g.,} ``close friends are those co-located outside of work''). Figure \ref{fig:realityb} reports the ground-truth social network, self-reported from a user survey, accurately reconstructed by the inferred network.

While the discovery that friends meet or call after work is not particularly surprising, this demonstrates the integration approach of these modalities for the simple edge prediction task. The principle components measure to infer the network edges also incorporates domain knowledge of work schedules. Previous work shows that human mobility in urban environments is highly periodic between a small set of locations (\emph{e.g.,} home and work) \cite{Eagle2009}. Therefore, incorporating these periodicities explicitly is a key aspect of this domain.

\begin{figure}
	\centering
	\subfigure[A network inferred from a university email corpus]{\label{fig:email1}\includegraphics[width=.75\columnwidth]{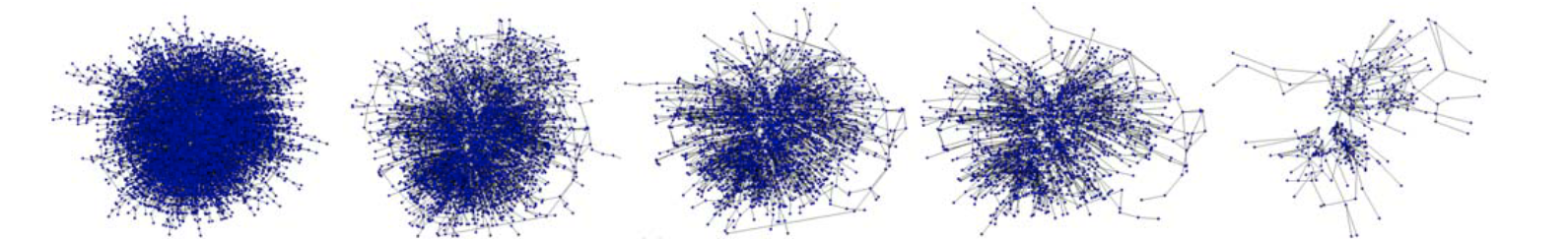}}
	\subfigure[A network inferred from the Enron email corpus]{\label{fig:email2}\includegraphics[width=.75\columnwidth]{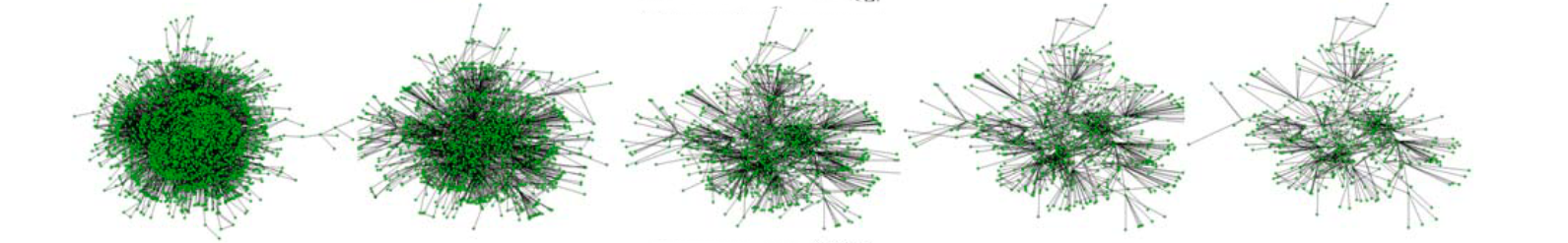}}
\caption{Two networks inferred over varying network thresholds $\tau$, creating sparser networks from left to right as the threshold criteria becomes more strict (\emph{e.g.,} more than $\tau$ email interactions for an edge between individuals $v_i$ and $v_j$). (From \citet{DeChoudhury:2010:IRS:1772690.1772722}).}
\label{fig:vary_threshold_network}
\end{figure}

\citet{DeChoudhury:2010:IRS:1772690.1772722} revisit the discussion of setting similarity threshold $\tau$ for an interaction measure calculated on data. Even under a fixed interaction measure, varying $\tau$ realizes many possible networks. In Figure \ref{fig:vary_threshold_network}, from left to right the number of required emails exchanged increases in order to define an edge, thus reducing the resulting network density. Any predictive task on this network balances novelty against the task difficulty: at a low threshold, a dense graph is realized and edge prediction may not perform better than random because the definition of the edge is simply noise. However, a very high threshold may infer a very sparse network, where edges are trivially easy to predict (but uninteresting).

The authors generate one-year aggregated networks in a university email dataset and the Enron email dataset (note that the time-scale is arbitrarily set). The authors tune the global threshold $\tau$ according to performance across several different classification tasks on sets of node-level features from each inferred network. These tasks include classification of node class (\emph{e.g.,} undergraduate, graduate, faculty, staff), gender, ``community'' (with class labels provided by stochastic block modeling, \cite{PhysRevLett.100.258701}), where each of these tasks may be independently of interest. If these node demographics and communities were separable as a set of behaviors at some ``natural'' threshold, this analysis would discover the threshold yielding the maximal classification accuracy. The authors also predict future communication activity using simple linear regression, reporting the accuracy at these same $\tau$. Each of these tasks yield a similar range of high performing $\tau$ thresholds, suggesting that classification and prediction agree across multiple views of the network.

\section{Conclusion}
\label{sec:conclusion}

This survey aims to provide a vocabulary and structure to the problem of inferring networks from data. Typically, this problem is addressed in data preprocessing, often with several \emph{artful} steps of parameter tuning or feature selection. We propose investigation of this problem using a more general and rigorous methodology for building networks appropriate for data science questions.

We survey several domains in order to illustrate their varying questions, challenges and how the nature of the data drives the methodological specializations in the areas. For example--with some simplification--we observe that gene regulatory networks are methodologically very mature, with a breadth of interaction measures appropriate for multivariate, \emph{matrix} data (\emph{e.g.,} microarray) including regression and graphical models. Climate networks and brain networks are mature in \emph{time series} interaction measures, including causal and frequency-based analysis, respectively. The problems in each of these areas are still exploratory, focusing on integrating and validating networks from different data (\emph{e.g.,} structural and functional brain networks) to develop data science tools downstream from these robust network models. Animal social networks inherit traditional \emph{direct observation} data in relatively simple formats, with instrumented sensing now becoming more feasible. This yields relatively simple network models over straightforward parameters (\emph{e.g.,} distance and persistence), with a focus on experimental design. Epidemiology historically studies observed infection data \emph{spreading} across a hidden contact network. Therefore, modeling these transmission functions is a key to this area. We hope that this \emph{data-driven} summary might help locate models and expertise on networks derived from different underlying data modalities.

Previous work often assumes that the objective of network inference is uncovering ``the network'' representation which is obscured by noise. Often in this context, the network inference method tries to reconstruct known ground-truth networks from data. In contrast, our work treats a network as a model to perform a particular \emph{task}, where we often cannot access the ground truth network, or assume its parametric form. Analogous to clustering for a classification \emph{task}, there are many possible clusterings which are only as valuable as the  accuracy improvement they provide of a downstream task (\emph{e.g.,} classification) or question. Conceptualizing network inference within the complete data science methodology--from data (to network) to task models for particular questions--focuses on a tighter coupling of network models constructed from data and task models. 

There are several methodologies across domains which use randomization, causality, and significance testing strategies to rigorously learn the network model under some assumptions. While these networks are appropriate according to their structural assumptions, they may not be the most informative for the question/task(s) of interest. Currently, no general, statistically rigorous methodology exists to learn and evaluate networks over particular task(s). Furthermore, there is little understanding of the criteria for network models and predictive models which would make them appropriate for this paired evaluation. We anticipate this will be an exciting area of future research and look forward to the followup publications.








\section{Acknowledgements}
This work was performed under the auspices of the U.S. Department of Energy by Lawrence Livermore National Laboratory under Contract DE-AC52-07NA27344 (LLNL-JRNL-703477), and with support of National Science Foundation grants  III-1514126 (Berger-Wolf), CNS-1248080 (Berger-Wolf), and IGERT CNS-1069311 (Brugere).
\begin{spacing}{.90}
\bibliographystyle{ACM-Reference-Format}
\renewcommand{\shortauthors}{Brugere, Gallagher and Berger-Wolf}
\bibliography{acmsmall-sample-bibfile_small}
\end{spacing}
\end{document}